\documentclass[10pt,final,journal,twoside]{IEEEtran} 
\usepackage[mathscr]{eucal}
\usepackage{ifpdf}
\usepackage{cite}
\usepackage{graphicx}
\usepackage[cmex10]{amsmath}
\usepackage{amssymb}
\usepackage{algorithmic}
\usepackage{units}
\usepackage{setspace}
\usepackage{algorithm}
\usepackage{array}
\usepackage[tight,footnotesize]{subfigure}
\usepackage{amsthm}
\usepackage{slashbox}
\usepackage{multirow}
\usepackage{enumerate}
\usepackage{color}
\usepackage{placeins}

\newtheorem{proposition}{Proposition}

\newtheorem{assumption}{Assumption}

\newtheorem{definition}{Definition}
\begin{document}
\title{A Non-Stationary Bandit-Learning Approach to Energy-Efficient Femto-Caching with Rateless-Coded Transmission} 
%
\author{
\IEEEauthorblockN{Setareh Maghsudi and Mihaela van der Schaar\\}
\thanks{S. Maghsudi is with the Electrical Engineering and Computer Science Department, Technical University of Berlin, 10623 Berlin, Germany (e-mail:maghsudi@tu-berlin.de). M. van der Schaar is with the Faculty of Engineering, University of California at Los Angeles, United States and the Faculty of Engineering, University of Cambridge, United Kingdom (e-mail:mihaela@ee.ucla.edu). A part of this paper appears at the 2019 IEEE Global Communications Conference \cite{MaghsudiGC}.}
}
\maketitle
%
\begin{abstract}
The ever-increasing demand for media streaming together with limited backhaul capacity renders developing efficient file-delivery methods imperative. One such method is femto-caching, which, despite its great potential, imposes several challenges such as efficient resource management. We study a resource allocation problem for joint caching and transmission in small cell networks, where the system operates in two consecutive phases: (i) cache placement, and (ii) joint file- and transmit power selection followed by broadcasting. We define the utility of every small base station in terms of the number of successful reconstructions per unit of transmission power. We then formulate the problem as to select a file from the cache together with a transmission power level for every broadcast round so that the accumulated utility over the horizon is maximized. The former problem boils down to a stochastic knapsack problem, and we cast the latter as a multi-armed bandit problem. We develop a solution to each problem and provide theoretical and numerical evaluations. In contrast to the state-of-the-art research, the proposed approach is especially suitable for networks with time-variant statistical properties. Moreover, it is applicable and operates well even when no initial information about the statistical characteristics of the random parameters such as file popularity and channel quality is available.
\end{abstract}
{\em Keywords}: Change detection, femto-caching, mortal and piece-wise stationary multi-armed bandits, rateless coding, stochastic knapsack problem. 
\section{Introduction}
\label{sec:Int}
To cope with the ever-increasing demand for mobile services, future wireless networks deploy dense small cells to underlay the legacy macro cellular networks \cite{Maghsudi17:TMS}, where small base stations (SBS) connect to the core network via a backhaul link. This concept takes advantage of low-power short-range base stations that potentially offload the macrocell traffic \cite{Maghsudi18:SCP}, improve the cell edge performance, and increase the uplink capacity. Moreover, transmission over short-length links enhances power efficiency and mitigates security concerns. However, these advantages come at some cost, and the system designers face several challenges to realize the concept of dense small cell networks. On the one hand, a massive number of devices sharing limited wireless resources in combination with a dense deployment of SBSs results in an excessive cost of information acquisition and computations, rendering centralized control mechanisms infeasible. On the other hand, in a dense small cell network, not all SBSs can access a power grid. Consequently, the SBSs store or harvest the required energy for providing wireless services. Naturally, the SBSs must consume energy efficiently to reduce the cost and the requirement of frequent batteries recharge, also to remain environment-friendly. 

Besides, the significant growth in the demand for media streaming in combination with limited backhaul capacity renders developing efficient file-delivery approaches imperative \cite{Schaar06:CBS}. The mobile data traffic caused by on-demand multimedia transmission exhibits the \textit{asynchronous content reuse} property \cite{Zink09:CYN}; that is, the users request a few popular files at different times, but at relatively small intervals. The concept of \textit{wireless caching} takes advantage of this property: instead of frequently fetching the popular files from the core network, different network entities store and re-transmit the files upon demand. The three major strategies for wireless caching are as follows: (i) femto-caching, (ii) D2D-caching, and (iii) coded-caching (or, coded-multicast) \cite{Ji16:WDD}. In femto-caching, SBSs save popular files to reduce the dependency on the high-speed backhaul. In D2D-caching, small devices store different files locally and provide each other with the files on-demand. Thus, in D2D-caching, the nearby devices share the burdens. The third method combines caching of files on the users' devices with the common multicast transmission of network-coded data. Despite the great potential to gain resource efficiency while improving the users' satisfaction levels, there are various challenges associated with caching. A challenging issue is the \textit{cache placement} problem, which refers to selecting a subset of available files to store. 
\subsection{Related Works}
\label{SubSec:Rel}
In the past few years, different methods and aspects of wireless caching have attracted a great deal of attention from the research community. In the following, we briefly review the cutting-edge research. In \cite{Ji16:WDD}, the authors provide an overview of the basics as well as the state-of-the-art of D2D-caching and compare various methods from a practical point of view. In \cite{Zhang16:ESP}, \textit{Zhang et al.} investigate the problem of joint scheduling and power allocation in D2D-caching. They decompose the problem into three sub-problems and solve each sub-problem in a centralized manner using convex optimization. D2D-caching is also the topic of \cite{Altieri14:ASG}. Assuming that the knowledge of the network's topology and files' popularity is available, the authors use stochastic geometry to design communication protocols. Probabilistic cache placement for D2D-caching is discussed in \cite{Kang14:MDD}. Similar to \cite{Altieri14:ASG}, the authors assume that a central controller knows the network's topology and files' popularity. Based on this assumption, they propose a solution to the formulated problem. In \cite{Dai12:CCW}, the authors consider the bandwidth allocation problem in proxy-caching. They propose a centralized solution based on auction theory that requires information about the files' popularity. Similarly, \cite{Zhang04:CBC} proposes a centralized heuristics for joint cache placement and server selection problem in proxy-caching. The developed solution necessitates knowledge of files' popularity. Physical layer caching in wireless ad-hoc networks is the focus of \cite{Liu16:ASL}. The authors propose a caching scheme that potentially enhances the capacity of wireless ad-hoc networks by changing the underlying topology. They also analyze asymptotic scaling laws. Caching methods based on network coding (coded-multicast) are investigated in \cite{Maddah14:FLC} and \cite{Shnaiwer15:NCC}, as examples. The former considers femto-caching in conjunction with opportunistic network coding and derives the optimal cache placement. The latter introduces the concept of \textit{global caching gain} against the conventional \textit{local caching gain}. The authors suggest a coded-caching scheme that exploits both gains. The focus of this paper is on \textit{femto-caching}; thus, in what follows, we confine our attention to research works that study this caching method. 

Reference \cite{Golrezaei12:FCW} introduces the femto-caching comprehensively. In \cite{Peng15:BAC}, the authors formulate the cache placement as an integer programming and solve it by developing a centralized method. Assuming that the files' popularity is known, \cite{Gregori16:WCC} studies offline D2D- and femto-caching. In both scenarios, they investigate the joint design of the transmission and caching policies by formulating a continuous-time optimization problem. The topic of \cite{Wang15:DFM} is dynamic femto-caching for mobile users in a network with a time-varying topology. It proposes two algorithms for sub-optimal cache placement. One of the algorithms is centralized. The other one, despite being decentralized, requires knowledge of the popularity profile, current network topology, and the cache of other SBSs. Resource allocation for a small cell caching system is investigated in \cite{Li15:EVP}, by using stochastic geometry to perform the analysis. In \cite{Wu19:DPT}, the authors investigate the trade-off between two crucial performance metrics of caching systems, namely, delivery delay and power consumption. They formulate a joint optimization problem involving power control, user association, and file placement. They solve the problem by decomposing it into two smaller problems. The focus of \cite{Liu16:EED} is on energy efficiency in cache-enabled small cell networks. They discuss different factors such as interference level, backhaul capacity, content popularity, and cache capacity concerning their impact on energy efficiency. They then investigate the required conditions under which caching contributes to energy efficiency.

While a great majority of research studies assume that SBSs are provided with some prior knowledge, for instance, about the files' popularity, there are only a few papers that do not rely on such assumptions when approaching different aspects of wireless caching. For example, \cite{Bharath16:ALB} defines the offloading time as the time overhead incurred due to the unavailability of the file requested by a typical user. Then, for an unknown popularity profile, a learning approach is proposed to minimize the offloading time. As another example, in \cite{Sengupta:LDC}, the authors study the learning of a distributed caching scheme in small cell networks. Reference \cite{Muller17:CAP} develops a learning-based cache placement scheme in small cell networks, where the SBS observes only the number of requests for the cached files (i.e., the cache hits), and not the requests for the other files (i.e., cache misses). Using the features of the requested content, the SBS learns which files to cache to maximize the cache hits in the future. In \cite{Li16:TAV}, the authors use an online approach to learn the files' popularity. Learning-based cache placement is also studied in \cite{Blasco14:LBO} and \cite{Muller16:SCW}. Although the aforementioned papers deal with unknown popularity, the developed methods do not include the power consumption, the time-variations in the files' popularity, and the possible gain of broadcast transmission. \textbf{Table \ref{Tb:SoA}} summarizes the contribution of some research papers.  
\begin{table*}[ht]
\caption{Wireless Femto-caching in Literature}
\label{Tb:SoA}
\begin{center}
\footnotesize
\begin{tabular}{|c|c|c|c|c|c|c|}
\hline
Reference            & Problem                   & Approach               & Information             & Concentration \\ 
\hline \hline
\cite{Peng15:BAC}    & Cache placement           & Convex optimization    & Local CSI, Popularity   & Centralized   \\ 
\hline
\cite{Poularak16:ECM}& Cache placement, Multicast& Randomized rounding    & Global CSI, Popularity  & Centralized   \\ 
\hline
\cite{Shim14:AFC}    & Cache placement, Overlapping helpers & Heuristics  & Popularity              & Centralized   \\ 
\hline
\cite{Gregori16:WCC} & Transmission and caching policy      & Optimization& Popularity              & Centralized   \\ 
\hline
\cite{Wang15:DFM}    & Cache placement                      & Heuristics  & Popularity, Topology    &  Centralized  \\ 
\hline
\cite{Li15:EVP}      & Resource allocation     & Stochastic geometry      & Popularity, Topology    & Distributed   \\ 
\hline
\cite{Yang14:EEW}    & Energy efficiency                    & Heuristics  & Popularity, Topology    & Centralized  \\ 
\hline
Our Work             & Cache placement, Resource allocation & MAB, Statistics      & None           & Distributed  \\
\hline
\end{tabular}
\end{center} 
\end{table*}  
\subsection{Our Contribution}
\label{SubSec:Cont}
A large body of cutting-edge research considers popularity as a known Zipf-like distribution. Such an assumption is, however, unrealistic due to the following reasons: (i) To a large extent, file popularity is dynamic and often vanishes over time; (ii) In general, the popularity of a file is an outcome of the decisions of a large population. Even if possible, such a variable is very costly to predict, especially due to massive production at a rapid pace. 
 
Moreover, the great majority of current research assumes that SBSs have access to an unlimited power resource; nevertheless, in dense small cell networks, many SBSs rely on a limited power supply. The scarcity of supply renders the power efficiency crucial. Therefore, it is essential to serve as many users as possible with a specific amount of power. This goal is difficult to achieve particularly in dense small cell networks, where due to a large number of users, acquiring global channel state information (CSI) at SBSs is remarkably expensive. In addition to efficiency, sophisticated power control has a significant impact on mitigating transmission impairments such as interference; nonetheless, most previous works exclude such physical-layer issues from the caching problem, disregarding the fact that the transmission policy of SBSs impacts the realized benefits of caching. More precisely, such exclusion might prevent developing optimal caching strategies: On the one hand, transmission power impacts the coverage area of every SBS and the number of successful downloads of a cached file. On the other hand, due to the dense deployment, even SBSs close to each other share the spectrum resources, so that the inter-cell interference in every small cell depends on the transmission power of the neighboring SBSs.     

We enhance the state-of-the-art as summarized in the following. Firstly, in comparison to the existing research, we employ a more general and realistic system model. In particular,
\begin{itemize}
\item We take the limited file and power storage capacities of SBSs into account.
\item In our setting, the SBSs do not have any prior information about the \textit{statistical characteristics} of the files' popularity or channel quality.
\item We consider a dynamic network model where the number of users associated with every SBS is a random variable.
\end{itemize}
Every SBS uses rateless coding to broadcast a file. When exploiting rateless codes, the receiver accumulates mutual information to decode the message \cite{Maghsudi12:DCR}. While the required time for decoding different blocks varies with channel conditions, rateless coding can guarantee zero outage probability, which makes it suitable for applications such as delay-constrained multimedia transmission over fading channels \cite{Maghsudi12:NCR}, \cite{Castura06:ORC}. We analyze the outage probability when rateless coding is applied. Afterward, we formulate the joint cache placement and power control problem in cache-enabled small cell networks. To deal with the NP-hardness of the formulated problem, we exploit the fact that caching systems usually work in two consecutive phases: (i) placement phase, and (ii) delivery phase \cite{Muller16:SCW}, \cite{Maddah14:FLC}. Therefore we decompose the challenge into two problems: (i) cache placement, and (ii) broadcast-file and -power selection, as described below:
\begin{itemize}         
\item To efficiently solve the online cache placement problem, we concentrate on the dynamics of file popularity, noting the fact that the popularity of every file is time-varying. We propose an algorithmic solution using a stochastic knapsack problem in combination with time-series analysis and statistical methods such as maximum likelihood estimation and change detection. We design the procedure and select the parameters carefully to achieve a good performance in terms of detecting variations in file popularity and fast reaction to changes.
\item In the delivery phase, to efficiently exploit the limited power supply at every SBS, we cast the selection of the optimal 
(file, transmit power) pair as a multi-armed bandit (MAB) problem with mortal arms. We adjust the algorithm and parameters so that it exhibits low regret, implying efficient performance in terms of quality of service (QoS) satisfaction and energy consumption. 
\end{itemize}
We further establish the applicability of our approach through extensive numerical analysis. 
\subsection{Paper Organization}
\label{SubSec:Org}
In Section \ref{Sec:SystemModel}, we introduce the network model. Afterward, in Section \ref{SubSec:Trans}, we design and analyze our rateless-coded transmission protocol. Section \ref{Sec:Cache} includes the description and analysis of the caching and files' popularity models. In Section \ref{Sec:Problem}, we formulate the joint cache placement and power control problem, which is then decomposed into two problems: (i) cache placement in the placement phase, and (ii) joint file- and transmit power selection in the delivery phase. In Section \ref{Sec:Mortal}, we solve the former problem by developing an algorithm based on the stochastic knapsack problem and some statistical methods. In Section \ref{Sec:Budget}, we cast and solve the latter problem as a multi-armed bandit game. We dedicate Section \ref{Sec:Numeric} to the numerical analysis. Section \ref{Sec:Conc} concludes the paper.
\subsection{Notation}
\label{SubSec:Not}
Throughout the paper, random variables are shown by uppercase letters while any realization is shown by lowercase. For example, $X$ and $x$ represent a random variable and its realization, respectively. The probability density or mass function (pdf) and the cumulative density function (cdf) of a random variable $X$ are shown by $f_{X}(x)$ and $F_{X}(x)$, respectively. By $\mathbb{P}\left[x\right]$ we denote the probability of some event $x$. Moreover, $\mathbb{E}\left[X\right]$ represents the expected value of $X$. We show a set and its cardinality by a unique letter, and distinguish them by using calligraphic and italic fonts, such as $\mathcal{A}$ and $A$, respectively. For any variable $a$, $\hat{a}$ represents its estimated value, which is calculated, for example, using maximum likelihood estimation. Matrices are shown by bold uppercase letters, for instance $\mathbf{A}$; thus $\left[\mathbf{A} \right]_{N\times M}$ denotes a matrix with $N$ rows and $M$ columns. Moreover, $\mathbf{A}_{l}$ denotes the $l$-th row of matrix $\mathbf{A}$. $[\mathbf{A}]_{l,m}$ stands for the element of matrix $\mathbf{A}$ located at $l$-th row and $m$-th column. Unit vectors are shown by bold lowercase letters, for example, $\textbf{a}$. The Hadamard product of two matrices $\mathbf{A}$ and $\mathbf{B}$ is shown as $\mathbf{A}{\circ}\mathbf{B}$. By $\mathbf{A}=\mathbf{B^{^{\circ\frac{1}{2}}}}$, we mean that matrix $\mathbf{A}$ is the element-wise square root of matrix $\mathbf{B}$. \textbf{Table \ref{Tb:Variables}} summarizes the variables that appear frequently in this paper.
\begin{table*}
\scriptsize
\caption{List of Frequently Used Variables  \footnotesize{(ordered as appear in the paper)}}
\begin{center}
\begin{tabular}{|c|c|}
\hline
Symbol                                         & Description                                                              \\ \hline \hline 
$\mathcal{M}$                                  & Set of SBSs (Deterministically deployed)                                         \\ \hline
$d$                                            & Small cell radius                                                                \\ \hline
$\mathcal{P}=\{p_{\min},..., p_{\max}\}$       & Set of power levels for SBSs                                                     \\ \hline
$p_{m}$                                        & Transmit power of SBS $m$                                                        \\ \hline
$\mathcal{N} \sim \textup{Poi}(\lambda)$       & Set of users (Poisson-distributed)                                               \\ \hline
$H_{nm} \sim \textup{Ray}(\frac{1}{\sqrt{2\beta_{nm}}})$ & Channel coefficient of link $m \to n$  (Rayleigh-distributed)          \\ \hline
$d_{nm}$                                       & Distance between SBS $m$ and user $n$                                            \\ \hline
$\mathcal{X}_{m} \subseteq \mathcal{N}$        & Set of users assigned to SBS $m$ (Distribution by (\ref{eq:Users}))              \\ \hline
$\mathcal{F}$                                  & Set of files to be potentially cached                                            \\ \hline
$L_{f}$                                        & Number of data blocks of file $f$                                                \\ \hline  
$L'_{f}$          & Minimum required number of successfully-received packets to recover the file with probability $\delta$        \\ \hline  
$\delta$          & Probability of successful file recovery if at least $L'_{f}$ packets are successfully received                \\ \hline 
$D_{m,f}$         & Maximum number of rateless-coded packets produced by SBS $m$ to broadcast file $f$                            \\ \hline
$R_{nm}$       & SINR of user $n$ assigned to SBS $m$ (Expression and distribution by (\ref{eq:SIRMin}) and (\ref{eq:SIRMinPdf})) \\ \hline
$O_{nm}$                                       & Outage probability of the link $m \to n$                                         \\ \hline
$u_{n,\min}$                                   & Minimum required transmission rate of user $n$ (Deterministic)                   \\ \hline
$P_{nm,f}^{(s)}$                               & Probability of successful recovery of file $f$ transmitted by SBS $m$ at user $n$\\ \hline
$W_{nm,f,p}$                                   & Time required by user $n$ to recover file $f$ transmitted by SBS $m$ at power $p$ \\ \hline
$W_{(X_{m}),f,p}$  & Maximum order statistic of $X_{m}$ random variables $W_{nm,f,p}$, $n \in \mathcal{X}_{m}$ (Distribution by 
(\ref{eq:MaxPdf}))                                                                                                                \\ \hline
$E_{m,f,p}$                                    & Total energy spent by SBS $m$ for broadcasting file $f$ at power $p$             \\ \hline
$C_{m}$; $S_{f}$                               & Cache size of SBS $m$; Size of file $f$ in bits                                  \\ \hline
$Q_{f,m,t}$        & Number of requests for file $f$ from SBS $m$ at time instant $t$ (Distribution by (\ref{eq:Requests}))       \\ \hline 
$\mathcal{A}_{\theta} \subseteq \mathcal{F}$   & Set of alive files at broadcast round $\theta$                                   \\ \hline
$f_{m,\theta}$; $p_{m,\theta}$                 & The broadcast file and transmit power of SBS $m$ at round $\theta$               \\ \hline
$E_{m,\theta}(f,p)$                            & Energy spent by SBS $m$ at round $\theta$                                        \\ \hline
$K_{m,\theta}(f,p)$                            & Number of users that recover the file successfully at broadcast round $\theta$   \\ \hline 
$g_{m,\theta}(f,p)$                            & Utility of SBS $m$ at round $\theta$                                             \\ \hline 
$\Gamma_{j-j'}$                                & Log-likelihood ratio for observations from time $j$ up to time $j'$              \\ \hline
$t_{a}$; $t_{c}$                               & Alarm time, Change time                                                          \\ \hline 
$\mathcal{I}_{m,\theta}=\mathcal{L}_{m,\theta} \otimes \mathcal{P}$ & Action set of SBS $m$ at broadcast round $\theta$           \\ \hline 
$i_{m,\theta}=(f_{m,\theta},p_{m,\theta})$     & Action of SBS $m$ at broadcast round $\theta$                                    \\ \hline
$i_{m,\theta}^{*}$; $g_{m,\theta}^{*}$         & Optimal action and Maximum achievable reward of SBS $m$ at broadcast round $\theta$\\ \hline
$d_{m,\theta}$                                 & Regret of SBS $m$ at broadcast round $\theta$                                    \\ \hline 
$\nu_{i,\theta}$                               & The index of arm $i$ at broadcast round $\theta$                                 \\ \hline 
$V_{i,\theta}$                                 & Number of times arm $i$ is played up to round $\theta$                           \\ \hline 
$\bar{g}_{i,\theta}$                           & Average reward of arm $i$ up to round $\theta$                                   \\ \hline
$q_{f,m,\theta}^{(T)}$                         & Number of requests for file $f$ from SBS $m$ at broadcast round $\theta$         \\ \hline
\end{tabular}
\end{center}
\label{Tb:Variables}
\end{table*}
\section{Network Model}
\label{Sec:SystemModel}
We study the downlink operation in an underlay cache-enabled small cell network with a \textit{set $\mathcal{M}$ of $M$ SBSs}. The service area consists of multiple grids with a radius of $d$. The deployment of SBSs is deterministic such that there exists only one SBS in every area with a radius of $d$. All SBSs are connected to a macro base station (MBS), which has access to the core network. Each SBS $m \in \mathcal{M}$ selects its transmission power from a finite \textit{set of power-levels} $\mathcal{P}$ including $P$ elements belonging to $\left[p_{\min},p_{\max} \right]$. As it is conventional in underlay networks, SBSs use the MBS's licensed channels if they are idle; thus, provided that the sensing is perfect, there is no interference incurred by- or to the MBS. The SBSs, however, might cause inter-cell interference to each other as they do not coordinate before transmissions. 

There exists a random \textit{set $\mathcal{N}$ of $N$ users} in the network. Users are quasi-static and their geographical arrangement follows a spatial homogeneous Poisson point process (HPPP) with density $\lambda$, which is unknown to the SBSs. We consider a frequency non-selective fading channel model. For a link connecting user $n \in \mathcal{N}$ and SBS $m \in \mathcal{M}$, we denote the random \textit{channel coefficient} by $H_{nm}$. The channel coefficient follows a Rayleigh distribution with unknown parameter $\frac{1}{\sqrt{2\beta_{nm}}}$. Therefore the channel gain $\left|H_{nm}\right|^{2}$ is exponentially-distributed with unknown parameter $\beta_{nm}$. The MBS, the SBSs, and the users do not have any prior knowledge, even statistical, about the channel quality.\footnote{Rayleigh fading is the most reasonable and widely-applied statistical model for propagation in an urban environment. Therefore, we use the Rayleigh distribution to develop the mathematical formulation and to enable the analysis. Note that it does not contradict our claim of the absence of prior statistical information at SBSs since we do not allow knowledge about the statistical characteristics of the distribution such as the expected value. It is worth noting that the analysis can be performed for any other fading model by following the same lines. Besides, the proposed joint cache placement and resource allocation method is applicable even if there is no mathematical model and analysis.}~Throughout the paper, we regard interference as noise. For the simplicity of notation and calculus, we neglect the large-scale fading (path loss), assuming that there is an inner power-control loop that compensates the path loss. This simplification does not affect our analysis, and the entire analysis can be performed in the same way also including the path loss.

Let $d_{nm}$ denote the distance between SBS $m \in \mathcal{M}$ and user $n \in \mathcal{N}$. Then user $n$ is associated with SBS $m$ if $d_{nm}=\textup{minimum}\{d_{n1},...,d_{nM}\}$. In words, every user is assigned to the nearest SBS. The distance-based user association is a widely applied association method (see \cite{Maghsudi17:DUA} and the references therein) since it is simple and practical. The \textit{set of users} assigned to every SBS $m \in \mathcal{M}$ is denoted by $\mathcal{X}_{m}$. Recall that (i) users are distributed according to an HPPP with density $\lambda$, and (ii) there is only one SBS in the area of radius $d$. Consequently, the \textit{number of users} assigned to SBS $m \in \mathcal{M}$ follows the distribution
\begin{equation}
\label{eq:Users}
f_{X_{m}}(x)=\frac{\left(\lambda \pi d^{2}\right)^{x}}{x!}e^{-\lambda \pi d^{2}}.
\end{equation}
It is worth mentioning that in ultra-dense small cell networks, a user might be located inside the intersection of the coverage areas of multiple SBSs. Hence, if the SBSs transmit the same file with specific coding constraints, the user benefits from multiple useful signals for rapid file reconstruction. However, we consider a scenario where each user is associated with only one SBS and receives recoverable signals only from that SBS. Therefore, each SBS decides on the files to cache independent of others. Also, note that the channel impairments such as interference can sometimes yield a packet delivery failure for some of the users assigned to an SBS.
\section{Transmission Model}
\label{SubSec:Trans}
Let $\mathcal{F}$ be the finite set of $F$ \textit{files to cache}. Every file $f \in \mathcal{F}$ consists of $L_{f}$ \textit{data blocks} of the same size.\footnote{Depending on the applied standard for wireless transmission, the size of each data block might change from a few Kilobytes (e.g., 1.5 Kilobytes for IPv4) and dozens of Kilobytes (e.g., 64 Kilobytes for IPv6). The exact size of the data block does not play a role in our proposed caching and transmission scheme since, if required, a data block of any size can be further segmented to blocks with a different size (often smaller) that is appropriate for rateless coding and transmission. Naturally, the receiver then re-integrates the blocks to each other as necessary to rebuild the original one.}~All SBSs apply rateless codes to transmit any file $f \in \mathcal{F}$, as described below. 

Every SBS $m \in \mathcal{M}$ first encodes the message at the packet-level using Luby-Transform (LT) rateless code \cite{Luby02:LT}. LT codes are the first practical realization of fountain codes that allow successful recovery of any message with very small overhead. In LT codes, every coded packet is the exclusive-or of $a$ neighbors. The number $a$ is called the \textit{degree}. It is selected according to some specific probability distribution. After being appended with a cyclic redundancy check (CRC) sequence, each LT-coded packet is encoded by a physical-layer channel code of rate $R$ bits/channel-use and is transmitted over the channel in a time slot. The receiver attempts to decode the packets at the physical layer first. If the CRC is incorrect, the corresponding packet is discarded. We assume that CRC bits are sufficient so that every error can be detected.\footnote{The overhead of CRC depends on the type being used, or, in other words, the number of CRC checksum bits. The checksum usually consists of 8, 16, or 32 bits in length and is appended to the message. In general, taking both overhead and error detection performance into account, CRC is considered superior to many other error detection codes such as parity check.}~The accumulation of correctly received packets is used by the decoder to recover the source file. Assume that the receiver desires to successfully reconstruct the message with probability $\delta$. Then it requires to accumulate (at least) any $L'_{f}=L_{f}+\nu_{f}(L_{f},\delta)$ LT-coded packets, provided that the number $a$ (i.e., the degree) is selected appropriately \cite{Luby02:LT}. Here, $\nu_{f}$ is the discrete random decoding overhead, rendering $L'_{f}$ a random variable as well; nonetheless, for point to point transmission, it can be deterministic. A common value is $\nu_{f}=0.05 L_{f}$ so that $L'_{f}=1.05 L_{f}$ \cite{Willkomm05:RLM}. The receiver then recovers the original message using, for instance, the belief propagation algorithm. It then informs the SBS via an acknowledgment (Ack) packet.\footnote{To decode the rateless-coded transmission, a receiver requires to possess channel state information. In our setting, each user receives signal only from one SBS, whereas the SBS transmits to multiple users; as such, it is plausible to assume that the receiver can obtain the CSI whereas the SBS incurs high cost for CSI acquisition. Also, some other articles such as \cite{Sun13:NCSI} suggest utilizing rateless codes to overcome the channel uncertainty at the transmitter. Moreover, some researchers address the problem of decoding rateless-coded data under imperfect CSI at the receiver. For example, in \cite{Venkiah08:RCQ}, the authors discuss rateless codes for quasi-static fading channels in conjunction with imperfect CSI at the receiver.} If the time and energy are unconstrained, rateless coding results in zero file-reconstruction failure also in fading channels. However, in a realistic scenario, either the energy is limited or the application is delay-sensitive. For example, in \cite{Castura06:ORC}, the authors apply rateless codes to wireless streaming with a sequence of playback deadlines. We also take such restrictions into account by assuming the following:  In every broadcast round, SBS $m \in \mathcal{M}$ generates \textbf{at most} $D_{m,f} \geq L'_{f}$ rateless-coded packets to broadcast any file $f \in \mathcal{F}$. Every SBS $m$ selects $D_{m,f}$ as a deterministic value. In essence, $D_{m,f}$ is a deadline for the users interested in file $f \in \mathcal{F}$ to accumulate $L'_{f}$ rateless-coded packets while SBS $m$ is broadcasting file $f$. The users with a higher channel quality are more likely to succeed in accumulating $L'_{f}$ rateless-coded packets out of $D_{m,f}$ transmitted ones. Naturally, $D_{m,f}$ shall be larger for larger files that consist of more packets. By selecting $D_{m,f}$, SBS $m$ balances the trade-off between the energy consumption for packet transmission on the one hand and the number of successful file reconstructions at the users' side on the other hand. Thus, the SBS stops transmitting as soon as one of the following events occurs: 
\begin{itemize}
\item All assigned users that have requested file $f$ recover it. 
\item $D_{m,f}$ data packets are transmitted. 
\end{itemize}
Let user $n \in \mathcal{N}$ be assigned to some SBS $m \in \mathcal{M}$, and the SBS transmits at some power level $p_{m} \in \mathcal{P}$. Then in the \textit{signal-to-interference-plus-noise ratio} (SINR) yields  
\begin{equation}
\label{eq:SIRMin}
R_{nm}(p_{m})=\frac{p_{m}h_{nm}^{2}}{\sum_{i \in \mathcal{M},i \neq m}p_{i}h^{2}_{ni}+p_{0}},
\end{equation}
where $p_{0}$ is the power of additive white Gaussian noise. Note that $R_{nm}$ attains a lower-bound in the worst-case scenario, where every SBS $j \in \mathcal{M}, j \neq m$, transmits with the maximum power, i.e., $p_{j}=p_{\max}$. The lower-bound, $R_{nm,\min}$, is useful when SBS $m$ is not aware of the transmit power of other SBSs. The pdf of $R_{nm}(p_{m})$, denoted by $f_{R}(r)$, is given by Proposition \ref{Pr:SirPdf}. 
\begin{proposition}
\label{Pr:SirPdf}
The pdf of SINR, $R_{nm}$, is given by
\begin{equation}
\label{eq:SIRMinPdf}
\begin{aligned}
f_{R}(r)=\\
\sum_{i \in \mathcal{M},i\neq m}&\beta'_{nm}\beta'_{ni}A_{ni}\exp(-\beta'_{nm}p_{0}r)\frac{(\beta'_{nm}r+\beta'_{ni})p_{0}+1}{(\beta'_{nm}r+\beta'_{ni})^{2}},
\end{aligned}
\end{equation}
where for all $i \in \mathcal{M}$ we define
\begin{equation}
\label{eq:SIRMinPdfT}
\beta'_{ni}=\frac{\beta_{ni}}{p_{i}}, 
\end{equation}
and
\begin{equation}
\label{eq:SIRMinPdfH}
A_{ni}=\prod_{l \in \mathcal{M}, l \neq i,m}\frac{\beta'_{nl}}{\beta'_{nl}-\beta'_{ni}}.
\end{equation}
\end{proposition}
\begin{IEEEproof}
See Appendix \ref{subSec:ProOne}.
\end{IEEEproof}
For a sufficiently long block length, the outage probability is a good approximation for the packet error probability. Let $u_{n,\min}$ denote the minimum transmission rate required by user $n \in \mathcal{N}$. The outage probability yields
\begin{equation}
\label{eq:Out}
O_{nm}(p)=\mathbb{P}\left[\log \left(1+R_{nm}(p)\right)<u_{n,\min}\right]=F_{R_{nm}}\left(e^{u_{n,\min}}-1 \right).
\end{equation}
where $F_{R_{nm}}(r)=\int_{0}^{r}f_{R}(r)dr$. In the general case, the solution of this integral at every desired point can be calculated numerically. For special cases, one can also calculate the closed-form solution. For example, for the interference-limited region where the noise power is negligible compared to the interference power, by using (\ref{eq:SIRMinPdf}), we can write (\ref{eq:Out}) as
\begin{equation}
\label{eq:OutOne}
O_{nm}(p) =
\sum_{i \in \mathcal{M}, i \neq m}A_{ni}\frac{\beta'_{ni}}{\beta'_{nm}}\frac{r_{n}}{r_{n}+\frac{\beta'_{ni}}{\beta'_{nm}}},
\end{equation}
where $r_{n}=e^{u_{n,\min}}-1$.\footnote{Depending on some factors such as the density of small base stations, small cell networks operate in different regions \cite{Yang18:DSC}.}~Naturally, the user experiences the maximum outage probability at $R_{nm,\min}$. A receiver can reconstruct a file $f \in \mathcal{F}$ with probability $\delta$ if it successfully receives at least $L'_{f}$ rateless-coded packets out of $D_{m,f}$ packets transmitted by SBS $m$. Hence, for a user $n \in \mathcal{N}$, the probability of successful decoding (file reconstruction) yields 
\begin{equation}
\label{eq:Time}
P_{nm,f,p}^{(s)}= \delta \left(1-\sum_{l=0}^{L'_{f}-1}\binom{D_{m,f}}{l}(1-O_{nm}(p))^{l}O_{nm}^{D_{m,f}-l}(p) \right).
\end{equation}
The quality of the wireless channel is random. Therefore, for each user, the required time until receiving enough packets for file reconstruction is a random variable. Moreover, when broadcasting a file $f \in \mathcal{F}$, each SBS $m \in \mathcal{M}$ transmits only a limited number of rateless-coded packets. Such restriction, in combination with random channel quality, renders the number of successful file reconstructions at the users' side as random. Consequently, the total power required by an SBS  to transmit a file is also a random variable. In the following sections, we analyze these random variables.
\subsection{Characterization of Energy Consumption}
\label{SubSec:CharPo}
Let $\mathcal{Q}_{f,m} \subseteq \mathcal{X}_{m}$ be the set of users that are associated with SBS $m \in \mathcal{M}$ and are interested in receiving a file $f \in \mathcal{F}$. Moreover, $W_{nm,f,p}$ denotes the number of rateless-coded packets transmitted by SBS $m$ with power $p$ until some user $n \in \mathcal{Q}_{f,m}$ accumulates $L'_{f}$ packets. Then $W_{nm,f,p}$ is a random variable following a negative binomial distribution, with the pdf 
\begin{equation}
\label{eq:TimePdf}
\begin{aligned}
f_{W_{nm,f,p}}(w)=\binom{w-1}{w-L'_{f}}\left(1-O_{nm}(p)\right)^{L'_{f}}O_{nm}^{w-L'_{f}}(p).
\end{aligned}
\end{equation}
Thus the cdf of $W_{nm,f,p}$ yields
\begin{equation}
\label{eq:TimeCdf}
\begin{aligned}
F_{W_{nm,f,p}}(w)&=\sum_{w=0}^{w} \binom{w-1}{w-L'_{f}}\left(1-O_{nm}(p) \right)^{L'_{f}}O_{nm}^{w-L'_{f}}(p)\\
&= \left(1-I_{O_{nm}(p)}\left (w-L'_{f}+1,L'_{f}\right)\right),
\end{aligned}
\end{equation}
where $I_{\{\cdot\}}$ is the regularized incomplete beta function. Assume that SBS $m \in \mathcal{M}$ continuously transmits rateless-coded packets until every user $n \in \mathcal{Q}_{f,m}$ receives $L'_{f}$ packets successfully, i.e., there is no limit as $D_{m,f}$. The number of rateless-coded packets required for that event is the maximum order statistic of the random variables $W_{nm,f,p}$, $n \in \mathcal{Q}_{f,m}$. For a deterministic value $A$, the cdf of the maximum order statistic of $A$ independent and non-identically distributed (i.ni.d) random variables, denoted by $W_{a}$, $a=1,..., A$, is given by \cite{David05:AGT}
\begin{equation}
\label{eq:MaxCdfDet}
F_{W_{(A)}}(w)=\sum_{\Gamma_{j}}\prod_{s=1}^{j}F_{rs}(w)\prod_{s=j+1}^{A}\left(1-F_{rs}(w)\right),
\end{equation}
where the summation $\Gamma_{j}$ extends over all permutations $\left(1,...,r_{A}\right)$ of $1,...,A$ random variables for which 
$1<\cdots<r_{j}$ and $r_{j}+1 <\cdots<A$. In our formulation, however, $Q_{f,m}$ (corresponding to $A$ here) is itself a \textit{random variable} following a distribution given by (\ref{eq:Users}); thus at every broadcast round, the \textit{largest reconstruction time}, denoted by $W_{(Q_{f,m})}$, has a pdf as  
\begin{equation}
\label{eq:MaxPdf}
\begin{aligned}
f_{W_{(Q_{f,m}),f,p}}(w)=\\
\sum_{q=0,1,...}\mathbb{P}\left[Q_{f,m}=q\right]&\left(F_{W_{(\cdot),f,p}}(w)-F_{W_{(\cdot),f,p}}(w-1)|Q_{f,m}=q\right),  
\end{aligned}
\end{equation}
where $F_{W_{(Q_{f,m}),f,p}}(w)$ is concluded by substituting (\ref{eq:TimeCdf}) in (\ref{eq:MaxCdfDet}). 

In practice, time and energy are constrained. Thus, every SBS $m \in \mathcal{M}$ transmits at most $D_{m,f}$ rateless-coded packets of any file $f \in \mathcal{F}$, unless all users $n \in \mathcal{Q}_{f,m}$ send the Ack signal to the SBS $m$ before $D_{m,f}$, implying that no further transmission is required. Using (\ref{eq:MaxPdf}), we calculate the probability of this event as
\begin{equation}
\label{eq:ReqTimeT}
\begin{aligned}
\mathbb{P}\left[W_{(Q_{f,m}),f,p}<D_{m,f}\right]&=\sum_{w=0}^{D_{m,f}-1}f_{W_{(Q_{f,m}),f,p}}(w)\\
&=F_{W_{(Q_{f,m}),f,p}}(D_{m,f}-1).
\end{aligned}
\end{equation}
Therefore, the \textit{duration of a broadcast round} $\theta=1,2,...$ in which an SBS $m \in \mathcal{M}$ transmits a file $f \in 
\mathcal{F}$ with power $p \in \mathcal{P}$ is a random variable defined as 
\begin{equation}
\label{eq:ReqTime}
\begin{aligned}
&T_{\theta}(m,f,p)\\
&=\begin{cases}
D_{m,f} & \text{with probability}~1-F_{W_{(Q_{f,m}),f,p}}(D_{m,f}-1) \\ 
w<D_{m,f} & \text{with probability}~f_{W_{(Q_{f,m}),f,p}}(w).
\end{cases}
\end{aligned}
\end{equation}
In (\ref{eq:ReqTime}), the first case corresponds to transmitting $D_{m,f}$ packets, whereas the second case occurs if the transmission ends sooner than that. By $T_{\theta=j}$, we denote the duration of some $\theta=j$. The \textit{total energy} spent by SBS $m \in \mathcal{M}$ is a random variable defined as 
\begin{equation}
\label{eq:ReqPow}
E_{m,f,p}= pT_{\theta}(m,f,p).
\end{equation}
Note that the maximum total energy is given by 
\begin{equation}
\label{eq:TPower}
E_{m,f,p}^{(\max)}=D_{m,f}p.
\end{equation}
%
\subsection{Characterization of the Number of File Reconstructions}
\label{SubSec:FileRecon}
Let $\mathcal{K}_{f,m,p} \subseteq \mathcal{Q}_{f,m}$ denote the set of interested users that recover the file successfully, i.e., 
\begin{equation}
\label{eq:ProbT}
\mathcal{K}_{f,m,p}=\left \{x \in \mathcal{Q}_{f,m}|W_{xm,f,p} \leq D_{m,f}\right\}.
\end{equation}
Since the channel quality is random, $K_{f,m,p}$ is a random variable. As discussed in Section \ref{SubSec:Trans}, one of the following events occurs:
\begin{itemize} 
\item All of the $Q_{f,m}$ requesting users reconstruct the broadcast file $f$ by some time $w<D_{m,f}$, or
\item A subset of users $\mathcal{K}_{f,m,p} \subseteq \mathcal{Q}_{f,m}$, where $K_{f,m,p} \leq Q_{f,m}$, reconstruct the file by the time 
$D_{m,f}$.  
\end{itemize}
The first event implies that the maximum order statistic of $Q_{f,m}$ random variables is equal to some $w<D_{m,f}$, which happens with probability $f_{W_{(Q_{f,m}),f,p}}(w)$. In the second event, the $K_{f,m,p}$-th order statistic of $Q_{f,m}$ random variables is equal to $D_{m,f}$, which happens with probability $f_{W_{(k),f,p}}(D_{m,f})$. Based on the argument above, $K_{f,m,p}$ can be characterized as 
\begin{equation} 
\label{eq:Reconstruct}
K_{f,m,p}=
\begin{cases}
Q_{f,m} & \text{with probability}~F_{W_{(Q_{f,m}),f,p}}(D_{m,f}) \\ 
k<Q_{f,m} & \text{with probability}~f_{W_{(k),f,p}}(D_{m,f})
\end{cases}
\end{equation}
%
\section{File Popularity and Caching Model}
\label{Sec:Cache}
Every SBS $m \in \mathcal{M}$ has a finite \textit{cache size}, denoted by $C_{m}$. Moreover, every file $f \in \mathcal{F}$ has a finite \textit{size} $S_{f}=L_{f}B$. Files are labeled in a way that $S_{1} \leq S_{2} \leq ... \leq S_{F}$. To gain access to a file that is not already stored in its cache, every SBS has to fetch it from the MBS. To avoid the energy-cost of fetching and to reduce the backhaul traffic, also to improve the users' satisfaction level through reducing the delay, the SBSs try to avoid repetitive fetching, by storing the most popular files in their cache. The vast majority of previous works models the file popularity by a Zipf-like distribution with a time-invariant skewness parameter. However, such a model stands in contrast to the regular experience where the popularity of most data files changes over time. For example, as a familiar pattern, the number of requests for a video clip tends to increase initially, remain static for a while and then decrease. Therefore, in our work, we model the file popularity as a piece-wise stationary random variable. Details follow.

For an SBS $m \in \mathcal{M}$ that serves $X_{m}$ users, the number of requests for a file $f \in \mathcal{F}$ follows a Poisson distribution with parameter $\mu_{f,t}X_{m}>0$;~that is, the total number of requests for each file is time-variant and depends on the number of users in the small cell. Every user can request multiple files. The SBSs have no prior information about the popularity of each file.\footnote{Traditionally, Poisson distribution describes the random \textit{arrival process}, e.g., user arrival or service demand in a telecommunication network. Therefore, here we use it to model the \textit{request arrival} for each file, although we consider the general setting in which the parameter of this distribution, i.e., the intensity is unknown. Moreover, it should be emphasized that based on the intensity, Poisson distribution produces values in $\mathcal{N}_{0}$.}~Let $Q_{f,m,t}$ be the \textit{total number of requests} for file $f \in \mathcal{F}$ submitted to SBS $m \in \mathcal{M}$ at time instant $t$. Then 
\begin{equation}
\label{eq:Requests}
\begin{aligned}
f_{Q_{f,m,t}}(q)=&\sum_{x=0,1,...}\mathbb{P}\left[X_{m}=x\right]\mathbb{P}\left[Q_{f,m,t}=q|X_{m}=x\right]\\ 
 =&\sum_{x=0,1,...}\frac{\left(\lambda \pi d^{2}\right)^{x}}{x!}\frac{\left(x \mu_{f,t} \right)^{q}}{q!} 
e^{-\lambda \pi d^{2}-x\mu_{f,t}},
\end{aligned}
\end{equation}  
where the equality follows by (\ref{eq:Users}) and the definition of Poisson distribution. 
\begin{assumption}
\label{as:PW}
The average file popularity is piece-wise constant so that $\mu_{f,t}$ is a step function that remains fixed over intervals of unknown length and suffers ruptures at change points. The values of $\mu_{f,t}$ before and after change times, as well as the change points, are unknown a priori. 
\end{assumption}
The time-variations in the files' popularity correspond to the notion of \textit{concept drift} in machine learning \cite{Minku10:TID}, which describes the changes in the statistical characteristics or distribution of the problem instances over time. For example, \cite{Tekin13:DOB} considers this notion. In essence, big data applications are often characterized by concept drift, in which trending topics change over time. Moreover, \cite{Tyson16:MAM} analyzes and illustrates some examples of concept drift in video streaming. Through data analysis, the authors show that the popularity of a video decreases by moving away from the posting date, or in other words, as the content's age increases. 

If $\psi$ and $\psi'$ are intensities before and after the change, we have $\left|\psi'-\psi \right|>C$, with $C$ being a sufficiently large constant. Note that the changes in the popularity can be continuous; however, by assuming a step-wise model, we neglect small changes in popularity and tune our algorithm to detect abrupt changes rather than small ones. In such a setting, the change detection performs better, since the tradeoff between a false alarm and misdetection can be balanced easier. Moreover, the complexity of caching reduces as well. More precisely, if even small changes are taken into account, the cache placement problem (knapsack problem) has to be solved unnecessarily very often, which results in excessive complexity. Based on this analytical model, in the following, we define the \textit{file type}. 
\begin{definition}[File Type]
\label{de:FiTy}
At any time instant $t$, a file $f \in \mathcal{F}$ is alive (i.e., still popular) if $\mu_{f,t}>\alpha_{m}$. Otherwise, it is dead; that is, it is not popular anymore.   
\end{definition}
We use $\mathcal{A}_{m,\theta} \subseteq \mathcal{F}$ to denote the set of files that are considered to be alive by SBS $m \in \mathcal{M}$, at the beginning of a broadcast round $\theta$. We assume that there always exists at least one alive file, i.e., $A_{m,\theta} \geq 1$ for all $\theta=1,2,...$. Naturally, as we later explain in Section \ref{Sec:Mortal}, every SBS selects the set of files for storage from the set of alive files, meaning that $\mathcal{L}_{m,\theta} \subseteq \mathcal{A}_{m,\theta}$. Thus, $\alpha_{m}$ is selected by SBS $m \in \mathcal{M}$ based on its cache size $C_{m}$. Intuitively, an SBS with larger storage capacity would have smaller $\alpha_{m}$, implying that it is able to store also some files with low popularity. 
\section{Problem Formulation}
\label{Sec:Problem}
Let $f_{m,\theta} \in \mathcal{A}_{\theta}$ and $p_{m,\theta} \in \mathcal{P}$ denote the broadcast file and transmit power of SBS $m$ at broadcast round $\theta$. Then, $E_{m,\theta}\left(f_{m,\theta},p_{m,\theta} \right):=E_{m,f_{m,\theta}, p_{m,\theta}}$ is the total energy spent during the broadcast round $\theta$. Moreover, let $K_{m,\theta}\left(f_{m,\theta}, p_{m,\theta} \right):=K_{m,f_{m,\theta},p_{m,\theta}}$ be the number of successful file deliveries. We define the utility of SBS $m$ at round $\theta$ as  
\begin{equation}
\label{eq:Rew}
g_{m,\theta}(f_{m,\theta},p_{m,\theta})=\frac{K_{m,\theta}\left(f_{m,\theta},p_{m,\theta}\right)}{E_{m,\theta}\left(f_{m,\theta},
p_{m,\theta} \right)},
\end{equation}
i.e., in terms of the number of successful reconstructions per energy consumption. Ideally, every SBS performs the cache placement and power allocation jointly. That is, it selects a file to cache, and at the \textit{same time} it assigns some power to broadcast that file (in future based on the users' request). Then the optimization problem of every SBS $m \in \mathcal{M}$ can be formulated as follows:
\begin{equation}
\label{eq:Prob}
\begin{matrix}
\underset{f_{m,\theta}\in \mathcal{L}_{m,\theta}\subseteq \mathcal{F}, p_{m,\theta} \in 
         \mathcal{P}}{\textup{maximize}}& \sum_{\theta=1,2,...}g_{m,\theta}(f_{m,\theta},p_{m,\theta}),
\end{matrix}
\end{equation}
subject to the following constraint
\begin{equation}
\label{eq:ConOne}
\sum_{f \in \mathcal{L}_{m,\theta}}S_{f}\leq C_{m}
\end{equation}
for all $\theta=1,2,...$. Solving the optimization problem in (\ref{eq:Prob}) is however not feasible since by the following reasons the objective function is not available: (i) The files' popularity, the network's structure (including number of users), and the CSI are not known a priori; (ii) Each SBS is only aware of its own transmit power level as well as the file being broadcast, and not those of other SBSs. In such a scenario, a natural solution would be to incorporate online learning methods that maximize the expected utility, that is,   
\begin{equation}
\label{eq:ExpProb}
\begin{matrix}
\underset{f_{m,\theta}\in \mathcal{L}_{m,\theta}\subseteq \mathcal{F}, p_{m,\theta} \in \mathcal{P}}{\textup{maximize}}& 
\mathbb{E}\left[\sum_{\theta=1,2,...}g_{m,\theta}(f_{m,\theta},p_{m,\theta})\right],
\end{matrix} 
\end{equation}
subject to (\ref{eq:ConOne}), where the expectation is taken with respect to random utility, as well as any other possible randomness in the decision-making strategy.

In general, efficient learning of the optimal decision involves finding a balance between gathering information about every possible strategy (\textit{exploration}) on one hand and accumulating utility (\textit{exploitation}) on the other hand. In other words, it is desired to learn as fast and efficient as possible, to minimize the cost of information shortage. In this sense, the problem formulated in (\ref{eq:ExpProb}) is inefficient to solve due to the following reasons: (i) Each (file, power level) pair, i.e., $(f,p)$, $f \in \mathcal{F}, p \in \mathcal{P}$ is defined as an action (strategy), yielding $FP$ actions in total; thus, from the computational point of view, such formulation imposes excessive complexity due to a large number of actions; (ii) To estimate the potential gain of each action, every file $f \in \mathcal{F}$ has to be fetched from the core network and transmitted at various power levels. It is evident that such exploration is very costly in terms of power, not only since fetching is power-consuming, but also since a substantial amount of limited power is wasted during exploration. Moreover, frequent fetching by densely deployed SBSs yields large backhaul traffic; (iii) The dynamic file popularity is not taken into account.  
  
\begin{figure*}[t]
\centering
\includegraphics[width=0.70\textwidth]{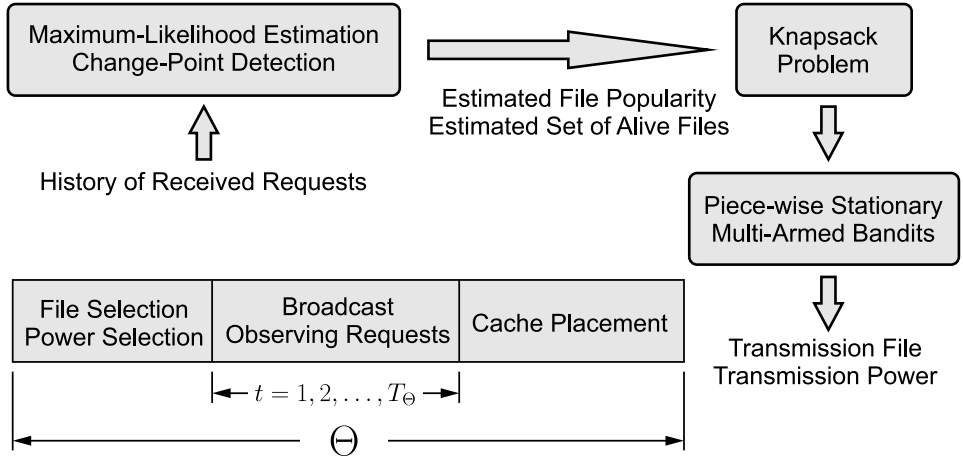}
\caption{The work flow and the timing structure for joint cache placement, broadcast file selection, and power control.}
\label{Fig:Flow}
\end{figure*}

Based on the discussion above and in agreement with many previous works such as \cite{Muller16:SCW} and \cite{Maddah14:FLC}, we consider the caching system to work in two consecutive phases: (i) cache placement phase, and (ii) delivery phase. Accordingly, we decompose the cache placement and power control problem into two sub-problems listed below, which are solved repeatedly and in parallel to converge to the optimal solution:
\begin{itemize}
\item \textbf{Cache Placement:} During every broadcast round $\theta=1,2,...$, every SBS $m \in \mathcal{M}$ observes the requests for every file $f \in \mathcal{F}$. Then, at the end of each broadcast round, the SBS updates its cache, if necessary. Later in Section \ref{Sec:Mortal}, we explain that the necessity is defied based on the variations in the files' popularity. 
\item \textbf{Joint Broadcast File Selection and Power Allocation:} At every broadcast round $\theta=1,2,...$, SBS $m \in 
\mathcal{M}$ selects a file $f_{m,\theta} \in \mathcal{L}_{m,\theta}$ to be transmitted, together with a power $p_{m,\theta} \in 
\mathcal{P}$ for broadcast.  
\end{itemize}
A summary of cache placement and delivery protocol is provided in \textbf{Algorithm \ref{alg:Protocol}}. The approach is also illustrated in \textbf{Fig. \ref{Fig:Flow}}. The two problems are described in detail and are solved in Section \ref{Sec:Mortal} and Section \ref{Sec:Budget}, respectively. In Section \ref{Sec:Deco}, we discuss the effect of decoupling the cache placement and power control problems on the performance.\footnote{Note that the frequency of updating the cache (if necessary at all) is not necessarily similar to that of broadcasting. In other words, an SBS can decide to update the cache in some specific intervals, for example, in every $K$ rounds of broadcasting. Our proposed algorithm is also applicable in this case.}
\begin{algorithm}
\caption{Summary of Caching and Transmission Protocol}
\label{alg:Protocol}
\small
\begin{algorithmic}[1]      
\FOR{$\theta=1,2,...$}
\STATE As described in \textbf{Algorithm \ref{alg:Cache}} (Section \ref{Sec:Mortal}), update the cache $\mathcal{L}_{m,\theta}$ if there is a change in files' popularity.
\STATE Select a file $f_{m,\theta} \in \mathcal{L}_{m,\theta}$ together with a transmission power level $p_{m,\theta} \in \mathcal{P}$ for broadcast, as described in \textbf{Algorithm \ref{alg:Multicast}} (Section \ref{Sec:Budget}). 
\ENDFOR 
\end{algorithmic}
\end{algorithm}
\section{Cache Placement}
\label{Sec:Mortal}
As described in Section \ref{Sec:Problem}, we perform the cache placement based on file popularity, or in other words, the requests. By (\ref{eq:ExpProb}), at every broadcast round, a file is selected from the cache; this implies that the dynamic cache placement is performed, if necessary, at broadcast (delivery) rounds and \textit{not} at every single transmission round. Let $\mu_{f,\theta}$ be the popularity of file $f \in \mathcal{F}$ at the beginning of the broadcast round $\theta$. For an SBS $m \in \mathcal{M}$, the cache placement problem is stated as follows: Given the set of files $\mathcal{F}$, where each file $f \in \mathcal{F}$ has a size $S_{f}$ and a value $\mu_{f,\theta}$, desired is to decide whether to include the file in cache ($\mathcal{L}_{m,\theta}$) or not, so that the total size of selected files does not exceed the cache capacity $C_{m}$, while the total value is maximized. Note that at every cache updating round, the saved files are selected from the set of alive files. Such a strategy reduces the complexity without affecting the performance adversely. Formally, for $\theta=1,2,...$, the optimization problem yields
\begin{equation}
\label{eq:KSproblem}
\begin{matrix}
\textup{maximize}&\sum_{f \in \mathcal{L}_{m,\theta}}\mu_{f,\theta}x_{f}\\ 
\textup{s.t.}&\mathcal{L}_{m,\theta} \subseteq \mathcal{A}_{m,\theta},\\
&\sum_{f \in \mathcal{L}_{m,\theta}}S_{f}x_{f} \leq C_{m},\\
& x_{f} \in \{0,1\}.\\
\end{matrix}
\end{equation}
Considering each file as an item whose value is defined in terms of popularity, the cache placement problem is a \textit{0-1 knapsack problem}, which is a combinatorial optimization problem. Note that in formulating the knapsack problem, the value of every item (file) $f \in \mathcal{F}$ can be also defined in terms of the number of requests; Nonetheless, according to our model, the number of requests is \textit{stochastic}, which renders the knapsack problem stochastic; Hence, to simplify the problem, we use the intensity of requests for each file $f$, $\mu_{f,\theta}$, as the item's value, keeping in mind that $\mu_{f,\theta}$ is deterministic, despite changing over time.  

Being under intensive investigation, a variety of efficient algorithmic solutions are developed for the knapsack problem 
\cite{Maghsudi17:DUAEH}. The cache placement problem stated by (\ref{eq:KSproblem}) is however more challenging than the conventional knapsack problem. In particular, according to our system model (see Section \ref{Sec:Cache}), (i) the SBSs do not have any prior knowledge on files popularity indicator $\mu_{f,\theta}$, and (ii) files' popularity is time-varying so that (\ref{eq:KSproblem}) has to be solved sequentially. Against such difficulties, we notice that at every time $t=1,2,...$, every SBS $m \in \mathcal{M}$ receives the requests for every file $f \in \mathcal{F}$. This side-information that every SBS obtains at almost no cost can be used to estimate the file popularity; i.e., to predict how many users would attempt to recover the file if it is broadcast. Let $\hat{\mu}_{f,\theta-1}$ be the estimated value of $\mu_{f,\theta-1}$ at the end of broadcast round $\theta-1$.\footnote{It is worth noting that in some previous works, for example, in \cite{Muller16:SCW} and \cite{Muller17:CAP}, the authors assume the following: Every SBS observes the cache hits, i.e., the number of requests for the files in the cache, whereas the number of cache drops, i.e., the number of requests for the files that do not belong to the cache, is not observable. However, such an assumption is unnecessary in many cases, where the users submit the requests for files to the corresponding SBS, while not knowing the cache status of the SBS. Even in the setting where the requests are submitted to some entity other than the SBS, such information can be transferred to the SBS with negligible cost.}~To update the cache for the broadcast round $\theta$, the SBS uses the historical data. Hence the optimization problem (\ref{eq:KSproblem}) is restated as 
\begin{equation}
\label{eq:KSproblemEst}
\begin{matrix}
\textup{maximize}&\sum_{f \in \mathcal{L}_{m,\theta}}\hat{\mu}_{f,\theta-1}x_{f}\\ 
\textup{s.t.}&\mathcal{L}_{m,\theta} \subseteq \hat{\mathcal{A}}_{m,\theta-1},\\
&\sum_{f \in \mathcal{L}_{m,\theta}}S_{f}x_{f} \leq C_{m},\\
& x_{f} \in \{0,1\}.\\
\end{matrix}
\end{equation}
For cache placement, we use the SBS's observations in combination with some \textit{statistical tests}, as described below.
\subsection{Maximum Likelihood Estimation}
\label{SubSec:MLE}
Consider a time period $\theta$ of length $T$, with its time instances being labeled as $1,...,T$. Let $\mathbf{Q}_{f,m,1:T}=
\left(q_{f,m,1},...,q_{f,m,T}\right)$ be a set of samples of the popularity of file $f \in \mathcal{F}$, where $q_{f,m,t}$ shows the number of requests for file $f$ submitted to SBS $m$ at time $j=1,...,T$. Assume that the popularity is stationary during the time interval $\left[1,T\right]$. Then the \textit{maximum likelihood estimate} of $\mu_{f,\theta}$ yields
\begin{equation}
\label{eq:MLE}
\hat{\mu}_{f,\theta}=\frac{1}{T}\sum_{t=1}^{T}q_{f,m,t}.
\end{equation}
%
\subsection{Change Point Detection}
\label{SubSec:CPD}
As described before, for any file $f \in \mathcal{F}$, the number of requests $q_{f,m,t}$ is a random variable following a Poisson distribution with parameter $\mu_{f,t}$. Therefore, the sequence of requests $q_{f,m,t}$, $t=1,2,...$, are i.ni.d random variables. Assume that the value of the parameter is $\mu_{f,t}=\psi_{0}$ initially at some reference time $t$, and at some unknown point of time, $t_{c}$, its value changes to $\mu_{f,t_{c}}=\psi_{1}$. \textit{Change point detection} basically detects the change and estimates the change time. Generally, it is assumed that $\psi_{0}$ is known, while $\psi_{1}$ might be known or unknown; in the latter case, the change magnitude $\left|\psi_{1}-\psi_{0} \right|$ or basically $\psi_{1}$ can be estimated as well. Formally, after each sampling, change point detection tests the following hypotheses about the parameter $\mu$:
\begin{equation}
\label{eq:HT}
\begin{aligned}
H_{0} :& \hspace{5pt} \mu=\psi_{0} \\ 
H_{1} :& \begin{cases}
\hspace{5pt} \mu \neq \psi_{0}            & \text{if}\hspace{3pt}\psi_{1}\hspace{3pt}\text{is unknown} \\ 
\hspace{5pt} \mu = \psi_{1}               & \text{if}\hspace{3pt}\psi_{1}\hspace{3pt}\text{is known}\\
\hspace{5pt} \left|\mu-\psi_{0}\right|>C  & \text{if changes are larger than}\hspace{3pt}C
\end{cases}
\end{aligned}
\end{equation}
In this paper, we utilize the \textit{generalized likelihood ratio (GLR)} test for change detection \cite{Basseville93}, as described in the following.\footnote{In addition to the GLR test, there exist several other methods for change-point detection. This includes Page-Hinkley test (P-H) \cite{Page54:CIS}, cumulative sum control chart (CUSUM) \cite{Barnard59:CCS}, sequential probability ratio test (CPRT), and the like. In this paper, we opt to use the GLR test since it applies to a wide range of probability distributions for the random variable under analysis. Moreover, it is suitable when the parameter's value before the change point as well as the change magnitude are unknown.}~For unknown $\psi_{1}$, the log-likelihood ratio for the observations from time $j$ up to time $j'$ is given by
\begin{equation}
\label{eq:GLRGen}
\Gamma_{j-j'}(\psi_{1})=\sum_{t=j}^{j'}\ln \left(\frac{f(q_{f,m,t};\psi_{1})}{f(q_{f,m,t};\psi_{0})}\right),
\end{equation}
where $f(x;\mu)$ is the probability of $x$ which depends on the scalar parameter $\mu$. For Poisson distribution, (\ref{eq:GLRGen}) yields 
\begin{equation}
\label{eq:GLRPoi}
\Gamma_{j-j'}(\psi_{1})=\sum_{t=j}^{j'}\ln \left(\frac{e^{\psi_{0}}\psi_{0}^{q_{f,m,t}}}{e^{\psi_{1}}\psi_{1}^{q_{f,m,t}}}\right).
\end{equation}
In (\ref{eq:GLRPoi}), two unknown variables are present. The first one is the \textit{change time} and the second one is the 
\textit{change magnitude} (value of $\psi_{1}$). These values can be calculated by using the maximum likelihood estimation; that is, we perform the following double maximization \cite{Basseville93}:
\begin{equation}
\label{eq:DoublMax}
\begin{matrix}
\gamma_{j'}=\underset{1\leq j\leq j'}{\max}&\underset{\psi_{1}:\left|\psi_{1}-\psi_{0} \right|\geq C>0}{\sup}~\Gamma_{j-j'}(\psi_{1}).
\end{matrix}
\end{equation}
Then the \textit{alarm time}, $t_{a}$, is defined as  
\begin{equation}
\label{eq:ta}
t_{a}=\min \left \{j':\gamma_{j'} \geq h \right\},
\end{equation}
where $h>0$ is an appropriately-selected constant.\footnote{The constant $h$ addresses the tradeoff between false alarm and detection delay, and should be determined so that a balance is achieved.}~Then the change point, $t_{c}$, and the mean value after the change, $\psi_{1}$, are estimated as \cite{Basseville93}
\begin{equation}
\label{eq:ChangTime}
(\hat{t}_{c},\hat{\psi}_{1})=\underset{1 \leq j \leq t_{a}}{\arg\max}~\underset{\psi_{1}:\left|\psi_{1}-\psi_{0}\right|\geq C>0}{\sup}~
\Gamma_{j-j'}(\psi_{1}).
\end{equation}
%
\subsection{The Algorithm}
\label{SubSec:Alg}
During the initialization period of length $T$, the SBS gathers observations of the requests for each file $f \in \mathcal{F}$ as statistical samples. Using this sample set and the maximum likelihood estimation described in Section \ref{SubSec:MLE}, the SBS estimates an initial popularity for every file $f \in \mathcal{F}$, denoted by $\hat{\mu}_{f,\theta=0}$. The SBS then solves the knapsack problem (\ref{eq:KSproblemEst}) to initialize the cache for $\theta=1$. In the course of each broadcast round $\theta=1,2,...$, and at every transmission trial, the SBS observes the requests for every file, which serve as new samples of popularity.\footnote{Note that similar to the Ack signals, requests can be submitted to each SBS also while transmitting, for instance through a control channel.}~The new samples are used to detect any changes in $\hat{\mu}_{f,t}$, and to estimate the change magnitude (i.e., the value of $\hat{\mu}_{f,t}$ after the change), as explained in Section \ref{SubSec:CPD}. At the end of each broadcast round $\theta$, the final value of $\hat{\mu}_{f,t}$ represents $\hat{\mu}_{f,\theta}$; Formally, $\hat{\mu}_{f,\theta}=\hat{\mu}_{f,T_{\theta-1}}$. Afterward, if necessary (i.e., upon the occurrence of some change), the SBS again solves the knapsack problem (\ref{eq:KSproblemEst}) and updates the cache. \textbf{Algorithm \ref{alg:Cache}} provides a summary of the procedure. 
\begin{algorithm}
\caption{Cache Placement}
\label{alg:Cache}
\small
\begin{algorithmic}[1]      
\STATE Initialization:
       \begin{itemize}
			   \item Select the length of initial observation $T$.
			   \item For every file $f \in \mathcal{F}$, the request vector $\mathbf{Q}_{f,m,1:T}=(q_{f,m,1},...,q_{f,m,T})$ is 
				       initialized with zero.
	       \item Let the initial set of alive files $\mathcal{A}_{0}=\varnothing$. Moreover, let the total run time so far be $T_{p}=0$.
				 \item Let $q_{f,m,\theta}^{(T)}=0$ denote the total number of requests for each file $f \in \mathcal{F}$ during broadcast
				       round $\theta$. 
				 \item Select the threshold for the popularity index to determine alive files, i.e., $\alpha_{m}$. 			
			 \end{itemize}
\FOR {$t=1,...,T$}
     \STATE For every file $f \in \mathcal{F}$, observe the number of requests at time $t$, i.e., $q_{f,m,t}$.
		 \STATE Update the total number of requests for every file $f \in \mathcal{F}$ as 
		$q_{f,m,0}^{(T)}=q_{f,m,0}^{(T)}+q_{f,m,t}$.			
\ENDFOR
\FOR {$f \in \mathcal{F}$}
		 \STATE Calculate $\hat{\mu}_{f,\theta=0}$, by using the sample set $\mathbf{Q}_{f,m,1:T}$ and the maximum likelihood estimate 
		        (\ref{eq:MLE}). 
		 \STATE If $\hat{\mu}_{f,\theta=0}>\alpha_{m}$, include file $f$ in $\hat{\mathcal{A}}_{0}$.   					
\ENDFOR
\STATE Solve the knapsack problem (\ref{eq:KSproblemEst}) by using the estimated values of $\hat{\mu}_{f,\theta=0}$, $f \in 
       \hat{\mathcal{A}}_{0}$, in order to initialize the cache ($\mathcal{L}_{m,1}$).
\FOR{broadcast rounds $\theta=1,2,3,...$}
     \STATE Initialization:
		      \begin{itemize}
					  \item Let $T_{\theta}$ denote the length of broadcast round $\theta$ (unknown a priori, see Algorithm \ref{alg:Multicast}). 
						\item For every file $f \in \mathcal{F}$: (i) The request vector $\mathbf{Q}_{f,m,1:T_{\theta}}$ is initialized with 
						      zero vector; (ii) Let the change detection flag $\mathbb{I}_{c,f}=0$; (iii) Let the change time $t_{c,f}=0$.
	          \item Let the estimated set of alive files $\hat{\mathcal{A}}_{\theta}=\varnothing$.				 
			 \end{itemize}
     \FOR{$t=1+T_{p},2+T_{p},...$} 
		      \FOR {$f \in \mathcal{F}$}
               \STATE Observe the number of requests at time $t$, $q_{f,m,t}$. 		 
							 \STATE Update the total number of requests for every file $f \in \mathcal{F}$ as $q_{f,m,\theta}^{(T)}=
							        q_{f,m,\theta}^{(T)}+q_{f,m,t}$.
					     \STATE Update the sample average as
								  \begin{equation}
								  \label{eq:Update}
								    \bar{q}_{f,m,t}=\frac{t-t_{c,f}-1}{t-t_{c,f}}\bar{q}_{f,m,t-1}+\frac{1}{t-t_{c,f}}q_{f,m,t}.
								  \end{equation}
					    \STATE Let $\psi_{0}=\bar{q}_{f,m,t}$ and perform the change point detection by using GLR test described by 
							       (\ref{eq:DoublMax}) and (\ref{eq:ta}).
					    \IF{a change is detected}
								  \STATE Find $\hat{\psi}_{1}$, $\hat{t}_{c}$ using (\ref{eq:ChangTime}).		
									\STATE Let $\hat{\mu}_{f,t}=\hat{\psi}_{1}$ and $t_{c,f}=\hat{t}_{c}$. 	
									\STATE Let $\bar{q}_{f,m,t}=\hat{\mu}_{f,t}$ and $\mathbb{I}_{c,f}=1$.								
							\ELSE		
							    \STATE Let $\hat{\mu}_{f,t}=\bar{q}_{f,m,t}$. 							   
							\ENDIF	
          \ENDFOR
					\STATE $T_{\theta}=T_{\theta}+1$.
		\ENDFOR
		\STATE For $f\in \mathcal{F}$, let $\hat{\mu}_{f,\theta}=\hat{\mu}_{f,t}$.
		\IF{$\exists \hspace{1pt} f \in \mathcal{F}:\mathbb{I}_{c,f}=1$}
      \STATE For $f\in \mathcal{F}$, if $\hat{\mu}_{f,t}>\alpha_{m}$, include $f$ in $\hat{\mathcal{A}}_{\theta}$. 
		  \STATE Solve the knapsack problem (\ref{eq:KSproblemEst}) by using $\hat{\mu}_{f,\theta}$, $f \in 
			       \hat{\mathcal{A}}_{\theta}$, to update the cache (find $\mathcal{L}_{m,\theta+1}$).
    \ELSE
		  \STATE $\mathcal{L}_{m,\theta+1}=\mathcal{L}_{m,\theta}$. 
		\ENDIF
		\STATE Let $T_{p}=T_{p}+T_{\theta}$.
\ENDFOR							
				
\end{algorithmic}
\end{algorithm}

It is worth mentioning that our proposed scheme to update the cache is event-triggered. More precisely, detecting a change in files' popularity results in an attempt to re-optimize the cache, i.e., to solve the knapsack problem (\ref{eq:KSproblemEst}). Based on the magnitude of the change(s), the re-optimization might yield an urge to fetch some new files from the core network via the backhaul link. Although an SBS incurs some cost for fetching new files, for example, to reimburse the resulted backhaul traffic, the situation is inevitable when the popularity of files is time-variant; Nevertheless, every SBS (or an MBS controlling the backhaul link) can reduce the burden of fetching by adapting its responsiveness to the changes in files' popularity. In other words, the SBS might decide to react only to a change whose magnitude is larger than some threshold value of $C$. Naturally, larger values of $C$ result in less sensitivity and hence the cache is updated less frequently. In the limit case, i.e., for very large values of $C$, our approach becomes similar to conventional caching schemes when the cache update is performed at specific time-intervals (time-organized rather than event-based). Finally, it should be noted that the knapsack problem (\ref{eq:KSproblemEst}) regards the size of every file as a cost factor; as fetching the large files naturally yields more backhaul traffic, one can conclude that the backhaul limitation is also taken into consideration in the optimization problem for cache placement.    
\section{Broadcast-File Selection and Power Control}
\label{Sec:Budget}
\textit{Multi-armed bandit} is a class of sequential optimization problems with incomplete information. The seminal problem involves an agent that is given a finite set of arms (actions, interchangeably), each producing some finite reward upon being pulled. Given no prior information, the agent selects arms sequentially, one at every round, and observes the reward of the played arm. Provided with this limited feedback, the agent aims at satisfying some optimality condition which is usually designed based on the problem's specific characteristics, such as the random nature of the reward generating process, duration of arms' availability, and so on. As a result of a lack of information, at each trial, the player may choose some inferior arm in terms of average reward, thus experiencing some regret. The \textit{regret} is the difference between the reward that would have been achieved had the agent selected the best arm and the actually achieved reward. 

In this paper, we model the joint file and power level selection as a multi-armed bandit problem, where every (file, power level) pair is regarded as one action. At every broadcast round $\theta=1,2,...$, a file $f_{m,\theta}$ is selected from the cache $\mathcal{L}_{m,\theta}$. The transmit power $p_{m,\theta}$ is selected from the set of power levels $\mathcal{P}$. Thus the agent has a set of actions $\mathcal{I}_{m,\theta}=\mathcal{L}_{m,\theta} \bigotimes \mathcal{P}$, which includes $I_{m,\theta}=L_{m,\theta}P$ elements. For simplicity, at every broadcast round $\theta$, we denote the \textit{action} as $i_{m,\theta}=\left(f_{m,\theta},p_{m,\theta} \right)$. The utility of this action is then given by (\ref{eq:Rew}). Ideally, every SBS desires to select the broadcast file and power level so as to maximize its expected accumulated utility through the entire horizon. Formally, 
\begin{equation}
\label{eq:AcRewMax}
\begin{matrix}
\underset{i_{m,\theta} \in \mathcal{I}_{m,\theta}}{\textup{maximize}}&\mathbb{E}\left[\sum_{\theta=1,2,...}g_{m,\theta}\left(i_{m,\theta}\right)\right].
\end{matrix} 
\end{equation}
However, due to the lack of prior information and also by the time-varying nature of the problem, (\ref{eq:AcRewMax}) is infeasible. As a result, every SBS would opt for a less ambitious goal, as we describe below. 

At every broadcast round $\theta=1,2,...$, we define the \textit{optimal action} as 
\begin{equation}
\label{eq:OptAct}
\begin{matrix}
i_{m,\theta}^{*}:=\left(f^{*}_{m,\theta}, p^{*}_{m,\theta}\right):=\underset{f \in \mathcal{L}_{m,\theta},p \in \mathcal{P}}{\arg\max}& 
g_{m,\theta}(f,p),
\end{matrix} 
\end{equation}
which results in a reward
\begin{equation}
\label{eq:OptRew}
g^{*}_{m,\theta}\left(i^{*}_{m,\theta}\right)=g^{*}_{m,\theta}\left(f^{*}_{m,\theta},p^{*}_{m,\theta}\right)
:=\underset{f \in \mathcal{L}_{m,\theta},p \in \mathcal{P}}{\max}g_{m,\theta}(f,p).
\end{equation}
The \textit{regret} of playing action $i_{m,\theta}$ at round $\theta$ is given by
\begin{equation}
\label{eq:RegT}
d_{m,\theta}=g^{*}_{m,\theta}\left(i^{*}_{m,\theta}\right)-g_{m,\theta}(i_{m,\theta}).
\end{equation}
Let $\mathcal{S}$ be the set of selection policies. Moreover, assume that SBS $m \in \mathcal{M}$ uses some policy $\sigma \in \mathcal{S}$ to select some action $i_{m,\theta}^{(\sigma)}$ at successive rounds $\theta=1,2,...$. Also, let $d_{m,\theta}^{(\sigma)}$ denote the regret caused by using policy $\sigma$ at round $\theta$. Then the goal of every SBS $m \in \mathcal{M}$ is to minimize its expected accumulated regret over the entire transmission horizon by using an appropriate decision-making strategy. Formally,
\begin{equation}
\label{eq:RegMin}
\begin{matrix}
\underset{\sigma \in \mathcal{S}}{\textup{minimize}}&\mathbb{E}_{\sigma}\left[\sum_{\theta=1,2,...}d_{m,\theta}^{(\sigma)}
\left(i_{m,\theta}^{(\sigma)}\right)\right].
\end{matrix} 
\end{equation}
Recall that after some broadcast rounds, the cache is updated based on the new estimations of the time-varying files' popularity. Consequently, the bandit's action set might vary over the broadcast rounds as well, since the cache, together with the set of power levels, determines the set of available actions; In other words, while some action $i_{m,\theta}=\left(f,p \right)$ might be included in the action set for some broadcast round $\theta$, there is a chance that it does not exist anymore in round $\theta'>\theta$, simply since the file $f$ is omitted from the cache as a result of losing its popularity. This type of bandit problem is referred to as multi-armed bandit with \textit{mortal arms}, i.e., a setting in which some arms might disappear and some other arms might appear with time. To solve the selection problem, we use the algorithm summarized in \textbf{Algorithm \ref{alg:Multicast}}, which is an adapted version of a policy that appears in \cite{Bnaya13:VMAB}. 

The algorithm is based on the \textit{upper-confidence bound policy} (UCB) \cite{Auer02:FTA}. Basically, in UCB, at every selection round, an \textit{index} is calculated for every action, which corresponds to an upper-bound of a confidence interval for the expected reward of that arm. At every round, the index of an arm depends on the number of rounds that specific arm is selected so far, together with its achieved rewards. Formally, assume that the action space is not time-varying and let $V_{i}$ be the number of times that an action $i \in \mathcal{I}$ is played so far, i.e.,
\begin{equation}
\label{eq:Num}
V_{i,\theta}=\sum_{s=1}^{\theta}\mathbb{I}_{\left\{i_{s}^{(\textup{UCB})}=i\right\}},
\end{equation}
where $\mathbb{I}_{\left\{x=y\right\}}$ is the indicator function that returns one if the condition $x=y$ holds and zero otherwise. Then, the average reward from pulling any action $i \in \mathcal{I}_{\theta}$ yields 
\begin{equation}
\label{eq:AvgRew}
\bar{g}_{i,\theta}=\frac{1}{V_{i,\theta}}\sum_{s=1}^{\theta}g_{i,s}\mathbb{I}_{\left\{i_{s}^{(\textup{UCB})}=i\right\}}.
\end{equation}
The UCB policy associates an index $\nu_{i,\theta}$ to each action $i \in \mathcal{I}$ at round $\theta$ \cite{Auer02:FTA}
\begin{equation}
\label{eq:Index}
\nu_{i,\theta}=\bar{g}_{i,\theta}+\beta \sqrt{\frac{\zeta \log(\theta)}{V_{i,\theta}}},
\end{equation}
with $\beta$ and $\zeta$ being appropriate constants. The second term on the right-hand side of (\ref{eq:Index}) is a padding function that guarantees enough exploration. The action with the largest index is selected to be played in the next round, i.e.,
\begin{equation}
\label{eq:SelectionRule}
i_{\theta+1}=\left(f_{\theta+1},p_{\theta+1} \right)=\underset{i \in \mathcal{I}_{\theta}}{\arg \max}~\nu_{i,\theta}.
\end{equation}
The time-variation of the action set and the mortality of arms render the described UCB policy insufficient; Hence, based on 
\cite{Bnaya13:VMAB}, we introduce the following adaptation into the procedure: 
\begin{itemize} 
\item If the cache is changed, play each new action (i.e., broadcast the new files with the available power levels); 
\item Set $V_{i,\theta}=1$ for all actions (old and new) but keep the average reward of the old files; 
\item Continue by the UCB procedure. 
\end{itemize}
The following proposition states the regret bound of the decision-making policy described in \textbf{Algorithm \ref{alg:Multicast}}.
\begin{proposition}
\label{th:Regret}
The expected regret $R_{\theta}$ of \textbf{Algorithm \ref{alg:Multicast}} is $O\left(B\log(\theta)\right)$, where $B$ is the number of times a new file has appeared in the cache during the broadcast rounds $1,...,\theta$.
\end{proposition}
\begin{IEEEproof}
See Appendix \ref{subSec:ProTwo}.
\end{IEEEproof}
\begin{algorithm}
\caption{Broadcast File Selection and Power Control}
\label{alg:Multicast}
\small
\begin{algorithmic}[1]  
\STATE Recall that $q_{f,m,\theta-1}^{(T)}$ denotes the total number of requests for each file $f \in \mathcal{F}$ during broadcast 
				       round $\theta-1$.
\STATE Recall that $\mathcal{L}_{m,\theta}$ is the cache before round $\theta$ starts, which is returned by 
       Algorithm \ref{alg:Cache}.						
\FOR{$\theta=1,2,3,...$}
  \STATE Let $T_{\theta}=0$ denote the length of broadcast round $\theta$ in terms of transmission trials (unknown a priori). 
  \FORALL{$i \in \mathcal{I}_{m,\theta}$}
    \STATE Calculate the file index, $\nu_{i,m,\theta-1}$, using (\ref{eq:Index}). 
  \ENDFOR
  \STATE Let $i_{m,\theta}=\left(f_{m,\theta},p_{m,\theta} \right)=\underset{i \in \mathcal{I}_{m,\theta}}{\arg \max}~\nu_{i,m,\theta-1}$.
 \REPEAT
  \STATE Send a rateless-coded packet.
  \STATE If any Ack signal is received, increase the number of successful file recoveries by one; i.e., 
	       let $K_{m,\theta}=K_{m,\theta}+1$.
	\STATE Let $T_{\theta}=T_{\theta}+1$.
  \UNTIL{$K_{m,\theta}=q_{f,m,\theta-1}^{(T)}$ (All interested users recover the file.) or $T_{\theta}=D_{m,f_{m,\theta}}$ 
        (Maximum possible number of rateless-coded packets are transmitted.).}
  \STATE Calculate the total energy spent during the current broadcast round by (\ref{eq:ReqPow}). 
	\STATE Calculate the total gain of the selection using (\ref{eq:Rew}).  
 \ENDFOR
\end{algorithmic}
\end{algorithm}
\subsection{Discussion on Optimality and Complexity}
\label{Sec:Deco}
As the final remark, we discuss the effect of decoupling the cache placement and power control. As described in Section \ref{Sec:Problem}, to solve the joint cache placement and the pair (broadcast file, power level) selection, we decompose the cache placement and power control problems. A decoupling of the optimization of the cache placement and the delivery phases reduces the complexity significantly and boosts the power efficiency. Despite these benefits and although such an approach is used also by some previous works, such decomposition might raise some concerns with respect to the optimality of the solution in terms of utility. Naturally, approximation and probabilistic analysis introduce a possible reduction in optimality which cannot be avoided; however, the following proposition describes the conditions under which the decomposition does not affect the optimality. 
\begin{proposition}   
\label{pr:Optimality}
A decoupling of the optimization of cache placement and delivery phases does not introduce any adverse effect on the optimality in terms of utility if
\begin{equation}
\label{eq:Optimality}
\underset{f \in \mathcal{L}_{m,\theta}, p \in \mathcal{P}}{\max} \mathbb{E}\left [g_{m,\theta}(f,p)\right] \geq \underset{f \in \mathcal{F}-\mathcal{L}_{m,\theta}}{\max}\frac{\mathbb{E}\left[Q_{f,m,\theta} \right]}{p_{\min}L'_{f}}.
\end{equation}
\end{proposition}
\begin{IEEEproof}
See Appendix \ref{subSec:ProThree}.
\end{IEEEproof}
Computationally, the proposed scheme consists of three main blocks: (i) The knapsack problem (\ref{eq:KSproblemEst}), which is an integer programming that accepts a fully polynomial-time approximation scheme in the number of alive files ($A$) and the cache capacity $C$ \cite{Kellerer04:KP}; (ii) The change detection, which is performed through maximum likelihood estimation that has a polynomial complexity in terms of the samples \cite{Ciuonzo17:GRT}; and (iii) The multi-armed bandit problem, whose solution involves calculating the arms' indexes and finding the largest index at every round, which, similar to the other blocks, has linear complexity in the number of actions, i.e., (file, power) pairs \cite{Knuth98:ACP}. 
\section{Numerical Results}
\label{Sec:Numeric}
\begin{figure}
\centering
\includegraphics[width=.70\linewidth]{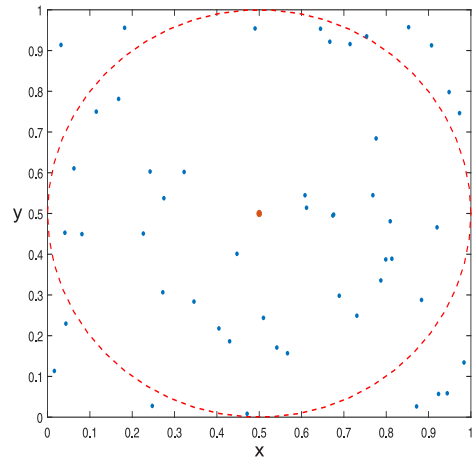}
\caption{An exemplary small cell model. The values on both axes are normalized by 100 meters.}
\label{Fig:Network}
\end{figure}

\begin{figure}
\centering
\includegraphics[width=.85\linewidth]{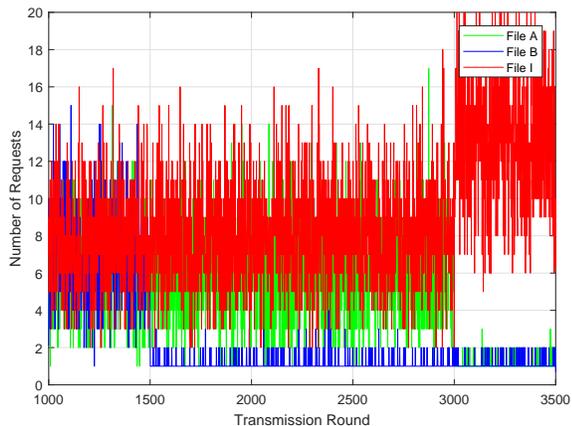}
\caption{Time-variations in the number of requests.}
\label{Fig:Change}
\end{figure}

We consider a small cell with one SBS where the users dynamically join and leave the network. At each time, the randomly-distributed users are drawn from a Poisson process with a density of $\lambda=38$. As an example, a snapshot is depicted in \textbf{Fig. \ref{Fig:Network}}. The SBS transmits using a power selected from the set $\mathcal{P}=\{1,2,4\}$. The SBS has a cache capacity of $C=15$. There are $F=10$ files to be cached potentially. There are two change points for popularity, $t=1.5\times 10^{3}$ and $t=3\times 10^{3}$. The files' characteristics, including the size, popularity density before the change and popularity density after the change points, are summarized in \textbf{Table \ref{Tb:Charcter}}. As an example, for File \textit{A}, File \textit{B} and File \textit{I}, i.e., two files whose popularity vanishes and one file whose popularity increases, the requests are shown in \textbf{Fig. \ref{Fig:Change}} as a function of time.  
\begin{table*}[!htbp]
\caption{Files Characteristics}
\label{Tb:Charcter}
\begin{center}
\footnotesize
\begin{tabular}{|c|c|c|c|c|c|c|c|c|c|c|}
\hline
File's Label                     &A    &B   &C  &D  &E  &F   &G  &H  &I   &J    \\ \hline 
\hline
Size                             &1    &1   &2  &5  &6  &3   &5  &4  &3   &7    \\ \hline 
Popularity Before Change         &5    &6   &3  &4  &6  &0.1 &1  &4  &7   &5    \\ \hline
Popularity After First Change    &5    &0.1 &3  &4  &6  &0.1 &1  &4  &7   &5    \\ \hline
Popularity After Second Change   &0.1  &0.1 &3  &4  &6  &0.1 &1  &4  &12  &5    \\ \hline
\end{tabular}
\end{center} 
\end{table*}  

\begin{figure*}[t]
\centering
\includegraphics[width=0.80\textwidth]{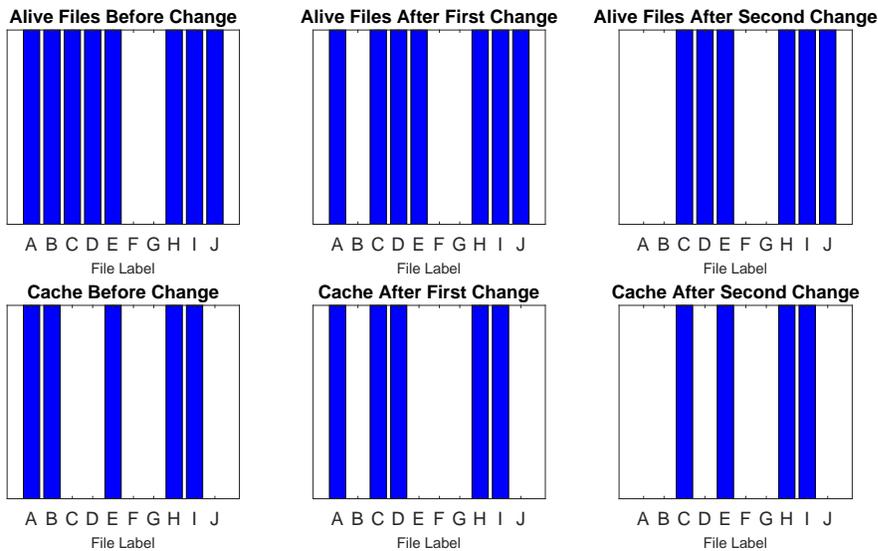}
\caption{Set of alive files and the cache, before and after the change points (popularity loss). In each sub-figure, the alive or selected files are marked by the blue line.}
\label{Fig:AlivCache}
\end{figure*}

The set of alive files, as well as the cache, are shown in \textbf{Fig. \ref{Fig:AlivCache}}, for both before and after the popularity change. From the figure, we can conclude the following: After a change occurs, the proposed cache placement method detects the change; Consequently, it replaces the file with the lost popularity with some popular file. By doing this, a cache update is triggered to adapt the cache to the occurred change. 

\begin{figure}[t]
\centering
\includegraphics[width=0.43\textwidth]{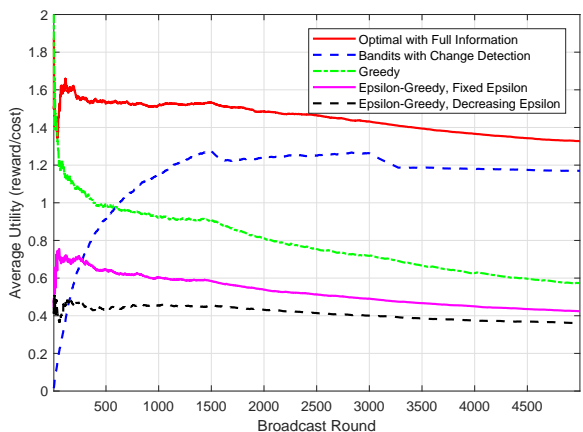}
\caption{The convergence of the average utility (in terms of reward/cost with unit 1/Joule) to the optimal value.}
\label{Fig:Bandit}
\end{figure}

\begin{figure}
\centering
\includegraphics[width=0.80\linewidth]{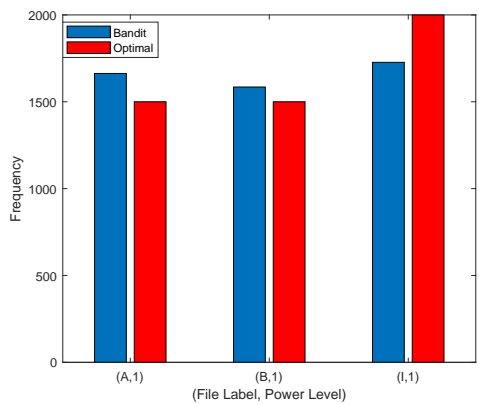}
\caption{The frequency of selecting each action by the proposed method compared to the optimal scheme. The actions with very low selection frequency are not shown.}
\label{Fig:Actions}
\end{figure}

\textbf{Fig. \ref{Fig:Bandit}} shows the utility of the proposed approach (mortal bandits with change-point detection), compared to the following methods: \textbf{(i) Optimal:} Given the full statistical information, the best option is selected using the exhaustive search; \textbf{(ii) Greedy:} After some initialization rounds which take place in the round-robin manner, the best arm so far is selected for the rest of the horizon; \textbf{(iii) Greedy with Fixed Exploration:} At every round, with probability $1-\epsilon$ the best arm so far, is selected and with probability $\epsilon$ an arm uniformly at random. The value of $\epsilon$ remains fixed over the entire horizon, which implies that the probability of exploration does not change; \textbf{(iv) Greedy with Decreasing Exploration:} This method is similar to the previous one, except that $\epsilon$ tends to zero with rate $1/t$, meaning that the probability of exploration decreases in time. From the figure, it is clear that the proposed approach detects the changes with an acceptable delay. Moreover, it adapts the set of alive files and following that, the cache. The algorithm then finds the optimal (file, power level) pair so that the average utility is almost equal to that of the optimal solution. Although the optimal solution performs slightly better than bandit in terms of the average utility, it imposes excessive computational complexity and cost for information acquisition. Moreover, \textbf{Fig. \ref{Fig:Actions}} shows the actions taken by the proposed algorithm (bandit) compared to those taken by the exhaustive search (optimal) approach. Most of the rounds in which the bandit algorithm does not select the optimal action belong to a short time interval right after the change, i.e., before the algorithm detects the change and adjusts the action. 

\begin{figure}
\centering
\includegraphics[width=0.76\linewidth]{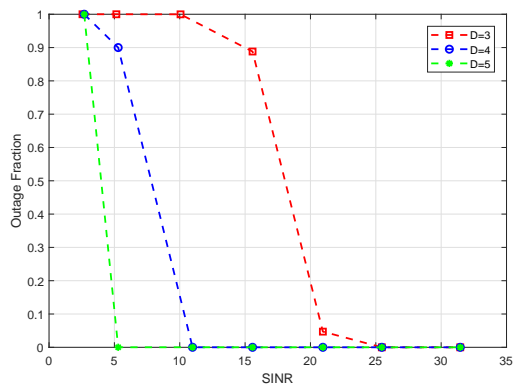}
\caption{The relation between SINR (the strength of desired and undesired signals), playback deadline (delay tolerance in seconds), and outage (QoS performance).}
\label{Fig:Rateless}
\end{figure}

Finally, we discuss the applicability and performance of rateless coding in delay-sensitive transmission, after SBS $m \in \mathcal{M}$ selects a file and a power level for transmission. Consider a large file with video content including 
$S_{f}$ segments, where each segment consists of $L_{f}$ data blocks. Due to the large size of the file $f$, for video transmission, the video-player does not wait for the entire video, i.e., all the $S_{f}\times L_{f}$ blocks to arrive before playback; i.e., each segment $i=1,2,...,S_{f}$ has some playback deadline $D_{i}=i\times D_{m,f}/L_{f}$. The blocks belonging to each segment are transmitted using rateless coding over the fading channel where we regard interference as noise. Therefore, for each segment $i=1,2,...,S_{f}$ at least $L'_{f}$ packets shall be arrived by the deadline, otherwise an outage occurs. In \cite{Castura06:RCF} and \cite{Maghsudi15:CSE}, the authors show that the outage performance has a strong relationship with the SNR or SINR (or, the transmission power $p$ and the channel quality) as well as the playback deadline (or, indirectly $D_{m,f}$). More precisely, a transition phase can be observed after which the outage of rateless transmission reduces abruptly to zero. \textbf{Fig. \ref{Fig:Rateless}} illustrates this effect, where we simulate the outage probability for delay-sensitive transmission of a video content with $S_{f}=1000$ segments through a fading channel as a function of SINR. Each segment consists of $L_{f}=2$ data blocks. Selecting $D_{m,f}=1000D$ with $D>1$, any segment $i=1,...,S_{f}$ should be decoded correctly after at most $D_{i}=i \times D$ channel uses (transmissions rounds), otherwise an outage occurs. From the figure, one can conclude that for different levels of delay-tolerance, different signal strengths are necessary to ensure zero outage.  
\section{Summary and Conclusion}
\label{Sec:Conc}
We considered a femto-caching scenario in a dynamic small cell network and studied a joint cache placement, broadcast file selection, and power control problem. We developed a solution for the formulated problem based on combinatorial optimization and multi-armed bandit theory. Theoretical and numerical results show that, in contrast to the conventional strategies, the proposed approach performs well under uncertainty, i.e., when the information about several variables such as files' popularity and channel characteristics is not available a priori. Moreover, it successfully adapts to time-varying popularity. The exact level of performance improvement depends on several factors such as the speed of variations in the files' popularity. Future research directions include an extension of the model and solution to the game-theoretical case where every user is served by multiple SBSs, implying that the cache placement and transmission can be done more efficiently. 
\section{Appendix}
\label{Sec:Appendix}
\subsection{Proof of Proposition \ref{Pr:SirPdf}}
\label{subSec:ProOne}
Let user $n \in \mathcal{N}$ be associated to SBS $m \in \mathcal{M}$. There is one desired signal, say $X$, which is exponentially-distributed with parameter $\beta'_{nm}=\beta_{nm}/p_{m}$. Moreover, there are $M-1$ interference signals $i$, where $i \in \mathcal{M}-\{m\}$. Hence the total interference power yields $Y=\sum_{i \in \mathcal{M},i \neq m}^{M}Y_{ni}$. Since $Y_{ni}$ is an exponentially-distributed random variable with parameter $\beta'_{ni}=\beta_{ni}/p_{i}$, $Y$ is the sum of i.ni.d exponential random variables. The pdf of $Y$ is given by \cite{Yao90:OPA}
\begin{equation}
\label{eq:pdfY}
f_{Y}(y)=\sum_{i \in \mathcal{M},i\neq m}^{M}A_{ni}\beta'_{ni}\exp(-\beta'_{ni}y),
\end{equation}
with $A_{ni}$ given by (\ref{eq:SIRMinPdfH}). Let $Z=Y+p_{0}$. Then by using the relation $F(Z\leq z)=F(Y \leq z-p_{0})$, the pdf of $Z$ can be calculated as 
\begin{equation}
\label{eq:pdfZ} 
\begin{aligned}
f_{Z}(z)&=\frac{d}{dz}\int_{y=0}^{z-p_{0}}f_{Y}(y)dy\\ 
&=\sum_{i \in \mathcal{M},i\neq m}^{M}A_{ni}\beta'_{ni}\exp(-\beta'_{ni}(z-p_{0}))\mathbb{I}_{\left \{ z\geq p_{0}\right\}.} 
\end{aligned}
\end{equation}
For $R=\frac{X}{Z}$, we have $f_{R}(r)=\int_{0}^{\infty}zf_{X}(rz)f_{Z}(z)dz$. Therefore the proposition follows by (\ref{eq:pdfY}).
\subsection{Proof of Proposition \ref{th:Regret}}
\label{subSec:ProTwo}
We follow the same line as in \cite{Bnaya13:VMAB}. As long as the cache does not change, the algorithm behaves like the standard UCB policy, which has a regret equal to $O\left(\log(\theta)\right)$. Removing a file from the cache does not affect this regret bound. Including a new file in the cache is equivalent to introducing $P$ new actions, since each action is defined as a (file, power level) pair. However, all of these $P$ new actions remain in the action set as long as the file $f$ remains in the cache, and disappear afterward. For $\theta$ broadcast rounds, the maximum number of rounds a file remains in the cache is $\theta$; In other words, if a file remains in the cache for some duration $\theta' \leq \theta$, the incurred regret is $O\left(\log(\theta')\right) \leq O\left(\log(\theta)\right)$. Therefore, the regret is bounded as $O\left(B\log(\theta)\right)$.
\subsection{Proof of Proposition \ref{pr:Optimality}}
\label{subSec:ProThree}
By decoupling the cache placement and delivery phases, the set of files $\mathcal{F}$ is divided into two parts: (i) The files that are stored in the cache, i.e., all $f \in \mathcal{L}_{m,\theta}$; and (ii) The files that are not stored in the cache, i.e., all $f \in \mathcal{F}-\mathcal{L}_{m,\theta}$. Assuming that the distance between the estimated file's popularity $\hat{\mu}_{f}$ and the true file popularity $\mu_{f}$ tends to zero, solving the knapsack problem (\ref{eq:KSproblemEst}) returns the optimal cache placement. Also, by Proposition \ref{th:Regret}, the bandit strategy finds the most efficient pair (broadcast file, power level) that maximizes the expected utility in terms of (\ref{eq:Rew}). Therefore, a sub-optimality occurs only if one of the files that are not included in the cache produces an expected utility larger than that of the pair found by the MAB policy. Formally, to guarantee the optimality, one needs to have
\begin{equation}
\label{eq:optOne}
\underset{f \in \mathcal{L}_{m,\theta}, p \in \mathcal{P}}{\max} \mathbb{E}\left [g_{m,\theta}(f,p)\right] \geq \mathbb{E}\left [g_{m,\theta}(f,p)\right] 
\end{equation}
for all $f \in \mathcal{F}-\mathcal{L}_{m,\theta}$ and $p \in \mathcal{P}$. The right-hand side of (\ref{eq:optOne}) can be upper-bounded as 
\begin{equation}
\label{eq:optTwo}
\begin{aligned} 
&\begin{matrix} \mathbb{E}\left [g_{m,\theta}(f,p)\right],\forall f \in \mathcal{F}-\mathcal{L}_{m,\theta},p \in \mathcal{P} 
\end{matrix}\\
&\geq \underset{f \in \mathcal{F}-\mathcal{L}_{m,\theta}, p \in \mathcal{P}}{\max} \mathbb{E}\left [g_{m,\theta}(f,p)\right]
\geq \frac{\mathbb{E}\left [K_{m,\theta}(f_{\theta},p_{\theta}) \right]}{\mathbb{E}\left [E_{m,\theta}(f_{\theta},p_{\theta})\right]}\\ 
&\geq \frac{\mathbb{E}\left [Q_{f,m,\theta}\right]}{p_{\min}L'_{f}}
=\frac{\mu_{f,\theta}}{p_{\min}L'_{f}},
\end{aligned}
\end{equation}
where the second inequality follows from the approximation $\mathbb{E}\left [\frac{K}{E}\right]\approx \frac{\mathbb{E}\left [K\right]}{\mathbb{E}\left [E\right]}-\frac{\mathbb{C}\left[K,E\right]}{\left (\mathbb{E}\left [E\right]  \right )^{2}}+\frac{\mathbb{V}\left[E\right]\mathbb{E}\left [K\right]}{\left (\mathbb{E}\left [E\right]  \right )^{3}}$ with $\mathbb{C}$ and $\mathbb{V}$ denoting the covariance and variance respectively, together with the fact that $K_{m,\theta}$ and $E_{m,\theta}$ are independent random variables with positive expected values. The third inequality follows from the prerequisite of decoding a rateless-coded message (see Section \ref{SubSec:Trans}) and transmission with minimum power level. 
\bibliographystyle{IEEEbib}
\bibliography{TWC}
\end{document}